\documentclass[12pt]{article}
\usepackage{amssymb}
\usepackage{amsfonts}
\usepackage{amsthm}

\newcommand{\alg}[1]{\mbox{$\mathcal{#1}$}}
\newcommand{\hil}[1]{\mbox{$\mathsf{#1}$}}

\title{Entanglement and Open Systems in Algebraic Quantum Field Theory}
\author{Rob Clifton and Hans Halvorson\footnote{Department of Philosophy,
 1001 Cathedral of
Learning, University of
Pittsburgh, Pittsburgh, PA 15260, USA, emails: rclifton+@pitt.edu, 
hphst1+@pitt.edu.}}
\begin{document}
\maketitle 
\begin{abstract}
Entanglement has long been the subject of discussion by philosophers 
of quantum theory, and has recently come to play an essential role 
for physicists in their development of quantum information theory.  
In this paper we show how the formalism of algebraic quantum 
field theory (AQFT) provides a rigorous framework within which to 
analyze entanglement in the context of a fully relativistic formulation 
of quantum theory.  What emerges from the analysis 
are new practical and theoretical limitations on an 
experimenter's ability to perform operations on a field in one spacetime region 
that can disentangle its state from the state of the field in other 
spacelike-separated regions.  These limitations show just how deeply 
entrenched entanglement is in relativistic quantum field theory, and yield a 
fresh perspective on the ways in which the theory differs conceptually 
from both standard
nonrelativistic quantum theory and classical relativistic field 
theory.         
  \end{abstract} 

  \begin{quote}
``\ldots despite its conservative way of dealing with physical 
principles, algebraic QFT leads to a \emph{radical change of 
paradigm}.  Instead of the Newtonian view of a space-time filled with 
a material content one enters the reality of Leibniz created by 
relation (in particular inclusions) between `monads' ($\sim$ the 
hyperfinite type III$_{1}$ local von Neumann factors $\alg{A}(O)$ which 
as single algebras are nearly void of physical meaning)'' Schroer  
(1998, p. 302).
\end{quote}
   
 \section*{1. Introduction}
 
In \emph{PCT, Spin and Statistics, and All That}, Streater and Wightman 
claim that, as a consequence of the axioms of algebraic quantum 
field theory (AQFT), ``it is difficult 
to isolate a system described by fields from outside effects'' (1989, p. 
139). 
Haag makes a similar claim in \emph{Local Quantum Physics}: 
``From the previous chapters of this book it is evidently not obvious 
 how to achieve a division of the world into parts to which one can 
 assign individuality\ldots Instead we used a division according to 
 regions in space-time.  This leads in general to open systems'' (1992, 
p. 298).
By a field system these authors mean that portion of 
a quantum field within a specified bounded open region $O$ of 
spacetime, with its associated algebra of observables $\alg{A}(O)$ (constructed in 
the usual way, out of field operators smeared with test functions having 
support in the region).  The environment of a field system, 
so construed, is naturally taken to 
be the field in the region $O'$, the spacelike complement of $O$. 
But then the claims above appear, at first sight, puzzling.  After 
all, it 
is an axiom of AQFT that the observables in $\alg{A}(O')$
commute with those in $\alg{A}(O)$. And this implies --- indeed, is 
\emph{equivalent} to --- the 
assertion that standard von Neumann 
measurements performed in $O'$ \emph{cannot} have `outside effects' on the 
expectations of observables in $O$ (L\"{u}ders, 1951). What, then, could the above authors 
possibly mean by saying that the field in $O$ must be regarded as an open system?  
  
A similar puzzle is raised by a famous passage in which 
Einstein (1948) contrasts the picture of physical reality 
embodied in classical field 
theories with that which emerges when we try to take quantum theory to be 
complete:
\begin{quote}
  ``If one asks what is characteristic of the realm of physical ideas 
  independently of the quantum theory, then above all the following 
  attracts our attention: the concepts of physics refer to a real 
  external world, i.e., ideas are posited of things that claim a 
  ``real existence'' independent of the perceiving subject (bodies, 
  fields, etc.)\ldots it appears to be essential for this arrangement of the 
  things in physics that, at a specific time, these things 
  claim an existence independent of one another, 
 insofar as these things ``lie in different parts of space''.  Without 
 such an assumption of the mutually 
      independent existence (the ``being-thus'') of spatially distant 
      things, an assumption which originates in everyday thought, 
      physical thought in the sense familiar to us would not be 
      possible.  Nor does one see how physical laws could be 
      formulated and tested without such clean separation.\ldots   
      For the relative independence of spatially distant things ($A$ 
      and $B$), this idea is characteristic: an external influence on 
      $A$ has no \emph{immediate} effect on $B$; this is known as the 
      ``principle of local action,'' which is applied consistently in 
      field theory.  The complete suspension of this basic principle 
      would make impossible the idea of the existence of (quasi-)closed systems and, thereby, the establishment of empirically 
      testable laws in the sense familiar to us''  (\emph{ibid}, 
      pp. 321-2; Howard's (1989) translation).
      \end{quote}
      There is a strong temptation to read 
Einstein's `assumption of the mutually independent existence of 
spatially distant things' and his `principle of local action' as 
anticipating, respectively, 
the distinction between separability and locality, or between 
nonlocal `outcome-outcome' correlation and
`measurement-outcome' correlation, which some philosophers argue is 
crucial to unravelling the conceptual implications of Bell's 
theorem (see, e.g., Howard 1989).   However, even in nonrelativistic 
quantum theory, there is no question of any nonlocal 
\emph{measurement}-outcome 
correlation between distinct systems or degrees of freedom, whose observables 
are always represented as commuting.  Making the reasonable assumption that 
Einstein knew this quite well, what 
is it about taking quantum theory at face value that
he saw as a threat to securing the existence of physically 
closed systems?  

What makes quantum systems open 
for Einstein, as well as for Streater and Wightman, and Haag, 
is that they can occupy entangled states in which 
they sustain nonclassical EPR 
correlations with systems outside 
their light cones.  That is, while it is 
correct to read Einstein's discussion of the 
mutually independent existence of distant systems as an implicit 
critique of 
the way in which quantum theory often represents their joint state as 
entangled, we believe it must be the \emph{outcome-outcome} EPR correlations 
associated with entangled 
states that, in Einstein's view, pose a problem for the 
legitimate testing of the predictions of quantum theory.  One could 
certainly doubt 
whether EPR correlations really pose any 
methodological problem, or whether they
 truly require the existence of physical (or `causal') influences 
acting on a quantum system from outside.  But the analogy with 
open systems in thermodynamics that Einstein and the others seem to be 
invoking 
is not entirely misplaced.  

Consider the simplest 
toy universe consisting of two nonrelativistic quantum systems, 
represented by a tensor product of two-dimensional Hilbert spaces 
$\mathbb{C}_{A}^{2}\otimes\mathbb{C}_{B}^{2}$, where system $A$ is the `object' system, 
and $B$ its `environment'.  Let $x$ be any state vector for the 
composite system $A+B$, and $D_{A}(x)$ be the reduced density 
operator $x$ determines for system 
$A$.  Then the von Neumann entropy of $A$, 
$E_{A}(x)=-\mbox{Tr}(D_{A}(x)\ln D_{A}(x))$ ($=E_{B}(x)$), 
varies with the degree to which $A$ and $B$ are entangled.  
If $x$ is a product vector with no entanglement, $E_{A}(x)=0$, whereas, at the opposite extreme, 
$E_{A}(x)=\ln 2$ when $x$ is, say, a singlet or triplet state.  The 
more $A$ and $B$ are entangled, the more `disordered' $A$ 
becomes, because it will have more than one state available to it and 
$A$'s probabilities of occupying them will approach equality. 
In fact, exploiting an analogy to Carnot's heat cycle and the second law 
of thermodynamics --- that it is 
impossible to construct a \emph{perpetuum mobile} --- Popescu and Rohrlich (1997) 
have shown that the general principle that it is impossible to create
entanglement between pairs of systems by local operations on one 
member of each pair  implies that the von Neumann 
entropy of either member provides the uniquely correct 
measure of their 
entanglement when they are in a pure state. 
Changes in their degree of entanglement, and hence 
in the entropy of either system, can only come about
in the presence of a nontrivial interaction Hamiltonian between them.  
But the fact remains that there is an intimate connection between a 
system's entanglement with its environment and the extent to which the 
system should be thought of as physically closed.   

Returning to AQFT, Streater and Wightman, as well as
Haag, all intend to make a far stronger claim about quantum field systems 
--- a point that even 
applies to spacelike separated regions of a \emph{free} 
field, and might well have offended Einstein's physical sensibilities 
even
more.  The point is that quantum field systems
 are \emph{unavoidably} and \emph{intrinsically} open to entanglement.  
 Streater and Wightman's comment is made in reference to the 
 Reeh-Schlieder (1961) theorem, 
 a consequence of the general 
axioms of AQFT.  We shall show
 that this theorem entails severe \emph{practical} obstacles to isolating 
 field systems from entanglement with other field systems.  Haag's comment goes deeper, and  
 is related to the fact that the algebras associated with 
field systems localized in spacetime regions are in all known models 
of the axioms type III von Neumann algebras.  We shall show 
that this feature of the local algebras imposes
 a fundamental limitation on isolating field systems from entanglement even 
\emph{in principle}.
 
Think again of our toy 
nonrelativistic universe $A+B$, 
with Alice in possession of system $A$, and the
state $x$ 
entangled.  Although there 
are no operations that Alice can perform on system $A$ which 
will reduce its entropy, she can still try to destroy its entanglement with 
$B$ by performing a standard von Neumann measurement 
on $A$.  If $P_{\pm}$ are the eigenprojections of the 
observable she measures, and the initial density operator of $A+B$ is 
$D=P_{x}$, where $P_{x}$ is the projection onto the ray $x$ 
generates, then the post-measurement joint state of $A+B$ will be given 
by the new 
density operator 
\begin{equation} \label{eq:Alice1} 
D \rightarrow D' = (P_{+}\otimes I) P_{x}(P_{+}\otimes I)+(P_{-}\otimes 
I)P_{x}(P_{-}\otimes I). 
\end{equation}
Since the projections $P_{\pm}$ are one-dimensional, and $x$ is 
entangled, there are 
nonzero vectors $a_{x}^{\pm}\in \mathbb{C}_{A}^{2}$ and 
$b_{x}^{\pm}\in \mathbb{C}_{B}^{2}$ such that
$(P_{\pm}\otimes I)x= 
  a_{x}^{\pm}\otimes b_{x}^{\pm}$, and a straightforward calculation 
  reveals that $D'$ may be re-expressed as 
\begin{equation}  \label{eq:Alice2}
D' = \mbox{Tr}[(P_{+}\otimes I)P_{x}]P_{+}\otimes 
P_{b_{x}^{+}}+\mbox{Tr}[(P_{-}\otimes I)P_{x}]P_{-}\otimes 
P_{b_{x}^{-}}.
\end{equation}
Thus, regardless of the initial state $x$, or the degree to which 
it was 
entangled, $D'$ will always be a convex 
combination of product states, and there will no longer be any entanglement between 
$A$ and $B$.  
One might say that Alice's operation on $A$ 
has the effect of isolating $A$ from any further
EPR influences from $B$.  

Moreover, this result can be 
generalized.  Given any finite or infinite dimension for the Hilbert 
spaces $\hil{H}_{A}$ and $\hil{H}_{B}$, there is always an operation Alice can 
perform on $A$ that will destroy its entanglement with 
$B$ no matter what their initial state $D$ was, pure 
or mixed.
In fact, it suffices for Alice to measure any nondegenerate 
observable $A$ with a discrete spectrum (excluding $0$).  The final 
state $D'$ will then be a convex combination of product states, 
each of which is a product density operator obtained by `collapsing' 
$D$ using some particular eigenprojection of the measured 
observable.  (The fact that disentanglement of a state can always 
be achieved in this way does not conflict with the recently 
established result there can be no `universal disentangling machine', i.e., 
no \emph{unitary} evolution that maps an arbitrary 
$A+B$ state $D$ to an unentangled state with the same 
reduced density operators as $D$ (Mor 1998; Mor and Terno 
1999).  Also bear in mind that we have \emph{not} required that a 
successful disentangling process 
leave the states of the entangled subsystems unchanged.) 

The upshot is that if entanglement \emph{does} pose a methodological threat, 
it can at least be brought under control in nonrelativistic quantum 
theory.  Not so when we consider the analogous setup in quantum field theory,  with Alice 
in the vicinity of one region $A$, and $B$ any other spacelike-seperated 
field system.  We shall see that AQFT puts both practical and 
theoretical
limits on Alice's ability to destroy entanglement between her field system 
and $B$.  Again, while one could doubt whether this poses any real methodological 
problem for Alice (an issue to which we shall return later), 
we think it is ironic, considering Einstein's point of view,
that such limits should be forced upon us precisely when we make 
the transition to a fully \emph{relativistic} formulation of quantum 
theory.

We begin in Section 2. by reviewing the formalism of 
AQFT, the concept of entanglement between spacelike separated field 
systems, and the mathematical representation of an operation 
performed within a local spacetime region on a field system.  In Section 
3., 
we connect the Reeh-Schlieder theorem with the practical difficulties 
involved in guaranteeing that a field system is disentangled from 
other field systems.  The language of operations also 
turns out to be indispensible for clearing up some apparently 
paradoxical physical implications of 
the Reeh-Schlieder theorem that have been raised in the literature 
without being properly resolved.  
In Section 4., we discuss differences
between type III von Neumann algebras and the standard type I von 
Neumann algebras employed in nonrelativistic quantum theory, 
emphasizing the radical implications type III algebras 
have for the ignorance interpretation 
of mixtures and entanglement.  We end Section 4. by 
connecting the type III character of the 
algebra of a local field system with the inability, in principle, to 
perform local operations on the system that will destroy its entanglement 
with other spacelike separated systems.  We offer this result as one 
way to make precise the sense in which AQFT requires a radical change 
in paradigm --- a change that, regrettably, has passed virtually
unnoticed by philosophers of quantum theory. 

\section*{2. AQFT, Entanglement, and Local Operations}

 First, let us recall that an abstract
$C^{*}$-algebra is a Banach $^{*}$-algebra, where the 
involution and norm are related by $|A^{*}A|=|A|^{2}$.  Thus 
the algebra $\alg{B}(\hil{H})$ of all bounded operators on a Hilbert space 
$\hil{H}$ is a $C^{*}$-algebra, with $^{*}$ taken to be 
the adjoint operation, and $|\cdot|$ the 
standard operator norm.  
Moreover, any $^{*}$-\emph{sub}algebra of $\alg{B}(\hil{H})$ 
that is closed in the operator norm is a $C^{*}$-algebra, and, 
conversely, one can show that every abstract $C^{*}$-algebra has a concrete 
(faithful) representation as a norm-closed $^{*}$-subalgebra of 
$\alg{B}(\hil{H})$, for some appropriate Hilbert space $\hil{H}$ 
(Kadison and Ringrose (henceforth, KR) 1997, Remark 4.5.7). 
On 
the other hand, a   
von Neumann algebra is always taken to be a concrete collection 
of operators on some fixed Hilbert space $\hil{H}$.  For $F$ any 
set of operators on $\hil{H}$, let $F'$ denote the 
commutant of $F$, the set of all operators on 
$\hil{H}$ that commute 
with \emph{every} operator in $F$.  Observe that $F\subseteq F''$, 
that $F\subseteq G$ implies $G'\subseteq F'$, and (hence) that $A'=A'''$.  $\alg{R}$ is called a 
von Neumann algebra exactly when $\alg{R}$ is a      
$^{*}$-subalgebra of $\alg{B}(\hil{H})$ that contains the identity and 
satisfies $\alg{R}=\alg{R}''$.  This is equivalent, via von Neumann's 
famous double 
commutant theorem (KR 1997, Theorem 5.3.1), to the assertion that $\alg{R}$ is closed 
in the strong operator topology, where $Z_{n}\rightarrow Z$ strongly just in 
case $|(Z_{n}-Z)x|\rightarrow 0$ for all $x\in\hil{H}$.  If the 
sequence $\{Z_{n}\}\subseteq\alg{R}$ converges to $Z\in\alg{R}$ in 
norm, then since 
$|(Z_{n}-Z)x|\leq |Z_{n}-Z||x|$, the convergence is also strong, hence
every von 
Neumann algebra is also a 
$C^{*}$-algebra.   However, not every $C^{*}$-algebra of operators is a von Neumann algebra.  
For example, the $C^{*}$-algebra $\alg{C}$ of all compact operators on 
an infinite-dimensional Hilbert space $\hil{H}$ --- 
that is, the norm closure of the $^{*}$-subalgebra of all finite 
rank operators on $\hil{H}$ --- does 
\emph{not} contain the identity, nor does $\alg{C}$ satisfy $\alg{C}=\alg{C}''$.  
(Indeed, $\alg{C}''=\alg{B}(\hil{H})$, because only
 multiples of the identity commute with all finite-dimensional 
projections, and of course 
\emph{every} operator commutes with all multiples of the 
identity.)
Finally, let $S$ be any self-adjoint (i.e., $^{*}$-closed) set of 
operators in $\alg{B}(\hil{H})$.  Then $S'$ 
is a $^{*}$-algebra containing the identity, and both $S'$ 
($=S'''=(S')''$) and $S''$ 
($=(S')'=(S')'''=(S'')''$)  
are von Neumann algebras.  If we suppose there is some other von 
Neumann algebra $\alg{R}$ such that $S\subseteq \alg{R}$, then 
$\alg{R}'\subseteq S'$, which in turn entails 
 $S''\subseteq \alg{R}''=\alg{R}$.  Thus 
$S''$ is actually the smallest von Neumann algebra containing 
$S$, i.e., the von Neumann algebra that $S$ generates.  
For example, the von Neumann algebra generated by all finite rank 
operators is the whole of $\alg{B}(\hil{H})$.  
 
The basic mathematical object of AQFT on 
Minkowski spacetime $M$ 
is an association 
$O\longmapsto \alg{A}(O)$
between bounded open subsets
$O$ of $M$ and $C^{*}$-subalgebras $\alg{A}(O)$ of 
an abstract $C^{*}$-algebra $\alg{A}$ 
(Horuzhy 1988, Haag 1992).
  The motivation for
this association is that the self-adjoint elements of 
$\alg{A}(O)$ represent the physical magnitudes, or observables, of 
the field intrinsic to the region $O$.  
We shall see below how the elements of $\alg{A}(O)$ can also be used to 
represent mathematically the physical operations that can be performed 
within $O$, 
and often it is only this latter 
interpretation of $\alg{A}(O)$ that is emphasized (Haag 1992, p. 104).  
One naturally assumes
\begin{center}  
\emph{Isotony:}  If $O_{1}\subseteq O_{2}$, then
$\alg{A}(O_{1})\subseteq
\alg{A}(O_{2})$.
\end{center}
  As a consequence, the collection of all local algebras 
  $\alg{A}(O)$ defines a net whose limit points can be used to define 
  algebras associated with unbounded regions, and in particular 
  $\alg{A}(M)$, which is identified with $\alg{A}$ itself.  
  
  One of the leading ideas in the algebraic approach to fields is that 
  all of the 
  physics of a particular field theory is encoded in the structure of 
  its net of local algebras.  (In particular, while any given field 
  algebra on $M$ obtained via smearing will define 
  a unique net, the net underdetermines the field algebra; see 
  Borchers 1960.) 
  But there are some
   general assumptions about the net $\{\alg{A}(O):O\subseteq M\}$ 
  that all physically reasonable 
  field theories are held
   to satisfy.  
  First, one assumes    
    \begin{center} 
\emph{Microcausality:}  $\alg{A}(O')\subseteq \alg{A}(O)'$.   
\end{center}
One also assumes
 that there is a
faithful representation $\mathbf{x}\rightarrow \alpha _{\mathbf{x}}$ of
the spacetime translation group of
$M$ in the group of automorphisms of $\alg{A}$, satisfying 
\begin{center} 
\emph{Translation Covariance:}  $\alpha
_{\mathbf{x}}(\alg{A}(O))=\alg{A}(O+\mathbf{x})$.
\end{center}
\begin{center}  
\emph{Weak Additivity:} For any $O\subseteq M$, $\alg{A}$ is the
smallest
$C^{*}$-algebra containing $\bigcup _{\mathbf{x}\in M}\alg{A}(O+\mathbf{x})$.
\end{center}
Finally, one assumes that there is some irreducible representation of the net 
$\{\alg{A}(O):O\subseteq M\}$ in which these local algebras are 
identified with 
von Neumann algebras acting on a (nontrivial) Hilbert space $\hil{H}$, $\alg{A}$ 
is identified with a strongly dense subset of $\alg{B}(\hil{H})$, and the following condition 
holds 
\begin{quote}
\emph{Spectrum 
Condition}: The generator of spacetime translations, the 
energy-momentum of the field, has a spectrum confined to the forward 
light-cone.
\end{quote}
These last three conditions, and their role in the proof of the 
Reeh-Schlieder theorem (microcausality is not needed), 
are discussed at length in Halvorson (2000).  
We wish only to note here that while the spectrum condition itself 
only makes sense relative to a representation --- wherein one can speak, 
via Stone's theorem, of a generator of the spacetime translation 
group of $M$ (now concretely represented as a strongly continuous group of unitary 
operators $\{U_{\mathbf{x}}\}$ acting on $\hil{H}$) --- the requirement that the abstract net 
\emph{have} a representation satisfying the spectrum condition does 
not require that one actually \emph{pass} to such a representation to compute 
expectation values, cross-sections, etc.  Indeed, Haag and Kastler 
(1964) have argued that there is a precise sense in which all concrete 
representations of a net 
are physically equivalent, including representations 
with and without a translationally invariant vacuum state vector 
$\Omega$.  Since we are not concerned with 
that argument here, we shall henceforth take the `Haag-Araki' approach 
of assuming that all the local algebras $\{\alg{A}(O):O\subseteq M\}$
 are von Neumann algebras 
acting on some $\hil{H}$, with $\alg{A}''=\alg{B}(\hil{H})$, and there 
is 
a translationally invariant vacuum 
state $\Omega\in\hil{H}$.    

We turn next to the concept of a state of the field.  Generally, a 
physical state 
of a quantum system represented by some von Neumann algebra 
$\alg{R}\subseteq\alg{B}(\hil{H})$ is given 
by a normalized linear expectation functional $\tau$ on 
 $\alg{R}$ that is both positive and countably additive.  Positivity is 
the requirement that $\tau$ map any positive operator in $\alg{R}$ to a 
nonnegative expectation (a must, given that positive operators 
have nonnegative spectra), while countable additivity is the requirement that 
$\tau$ be additive over countable sums of mutually orthogonal 
projections in $\alg{R}$.  (There are 
also non-countably additive or `singular' states on $\alg{R}$ (KR 1997, p. 723), but 
whenever we use the term `state' we shall mean \emph{countably additive} state.)  
Every state on $\alg{R}$ extends to a  
state $\rho$ on $\alg{B}(\hil{H})$ which, in turn, can be represented by a 
density operator $D_{\rho}$ on $\hil{H}$ via the standard 
 formula $\rho(\cdot)=\mbox{Tr}(D_{\rho}\cdot)$ (KR 1997, p. 462).  
A pure state on $\alg{B}(\hil{H})$, i.e., one that is not a nontrivial convex combination or 
mixture 
of other states of $\alg{B}(\hil{H})$, is then represented by a vector 
$x\in\hil{H}$.  We shall always use the notation $\rho_{x}$ for the 
normalized state functional $(x,\cdot x)/|x|^{2}$ ($=\mbox{Tr}(P_{x}\cdot)$).  
If, furthermore, we consider the 
restriction $\rho_{x}|_{\alg{R}}$, the induced state on some von 
Neumann subalgebra $\alg{R}\subseteq\alg{B}(\hil{H})$, we cannot 
in general 
expect it to be pure on $\alg{R}$ as well.  For example, with $\hil{H}=\mathbb{C}_{A}^{2}\otimes 
\mathbb{C}_{B}^{2}$, $\alg{R}=\alg{B}(\mathbb{C}_{A}^{2})\otimes I$, and $x$ 
entangled, we know that the induced state $\rho_{x}|_{\alg{R}}$, represented 
by $D_{A}(x)\in\alg{B}(\mathbb{C}_{A}^{2})$, is 
\emph{always} mixed.  Similarly, one cannot expect that 
a pure state $\rho_{x}$ of the field algebra 
$\alg{A}''=\alg{B}(\hil{H})$ --- which supplies a maximal 
specification of the state of the 
field \emph{throughout} spacetime --- will have a restriction to a 
local algebra $\rho_{x}|_{\alg{A}(O)}$ that is itself pure.  In fact, 
we shall see later that the Reeh-Schlieder theorem entails that the vacuum 
state's restriction to any local algebra is always highly mixed.

There are two topologies on the state space of a von Neumann algebra 
$\alg{R}$ that we shall need to invoke.  One is the metric topology 
induced by the norm on linear functionals.  The norm of a state $\rho$ 
on $\alg{R}$ is defined by $\|\rho\|\equiv\sup\{|\rho(Z)|: 
Z=Z^{*}\in\alg{R},\ |Z|\leq 1\}$.  If two states, $\rho_{1}$ and 
$\rho_{2}$, are close to each other in norm, then they dictate close 
expectation values uniformly for \emph{all} observables.   In 
particular, if both 
$\rho_{1}$ and $\rho_{2}$ are vector states, i.e., they are induced by vectors 
$x_{1},x_{2}\in\hil{H}$ such that $\rho_{1}=\rho_{x_{1}}|_{\alg{R}}$ 
and $\rho_{2}=\rho_{x_{2}}|_{\alg{R}}$, then $|x_{1}-x_{2}|\rightarrow 
0$ implies $\|\rho_{1}-\rho_{2}\|\rightarrow 
0$.  (It is important not 
to conflate the terms `vector state' and `pure state', unless of 
course $\alg{R}=\alg{B}(\hil{H})$ itself.)  
More generally, whenever the trace norm distance between two density operators 
goes to zero, the norm distance between the states they induce on 
$\alg{R}$ goes to zero.  Since every state on $\alg{B}(\hil{H})$ is 
given by a density operator, which in turn can be decomposed as 
an infinite convex combination of one dimensional projections (with the 
infinite sum understood as trace norm convergence), it 
follows that every state on $\alg{R}\subseteq\alg{B}(\hil{H})$ is the norm limit of 
convex combinations of vectors states of $\alg{R}$ (cf. KR 1997, Thm. 
7.1.12).     
The other topology we shall invoke is the weak-$^{*}$ topology: 
a sequence or net of states $\{\rho_{n}\}$ on 
$\alg{R}$ weak-$^{*}$ converges 
to a state 
$\rho$ just in case $\rho_{n}(Z)\rightarrow\rho(Z)$ for all 
$Z\in\alg{R}$.  This convergence need not be uniform on all elements 
of $\alg{R}$, and is therefore weaker 
than the notion of approximation embodied by norm 
convergence.  As it happens, any state on $\alg{B}(\hil{H})$ that is 
the weak-$^{*}$ limit of a set of states is also their norm limit, but this 
is only true for type I von Neumann algebras
(Connes and St\o rmer, 1978, Cor. 9).

Next, we turn to defining entanglement in a field.  Fix a state $\rho$  on 
$\alg{B}(\hil{H})$, and
two mutually commuting subalgebras $\alg{R}_{A},\alg{R}_{B}\subseteq\alg{B}(\hil{H})$.
  To define what it means for $\rho$ to be 
entangled across the algebras, we need only consider the 
restriction $\rho|_{\alg{R}_{AB}}$ to the von Neumann algebra they generate, 
$\alg{R}_{AB}=[\alg{R}_{A}\cup\alg{R}_{B}]''$, and of course we need a 
definition that also applies when 
$\rho|_{\alg{R}_{AB}}$ is mixed.
A state $\omega$ on $\alg{R}_{AB}$ is called a product 
 state just in case there are states $\omega _{A}$ of
$\alg{R}_{A}$ and $\omega _{B}$ of $\alg{R}_{B}$ such that
$\omega (XY)=\omega _{A}(X)\omega _{B}(Y)$ for all 
$X\in\alg{R}_{A}$, $Y\in\alg{R}_{B}$.
Clearly, product states, or convex combinations of product states, 
possess only classical correlations.  Moreover, if one can even just 
\emph{approximate} a state with convex combinations of product 
states, its correlations do not significantly depart from 
those characteristic of a classical statistical theory.  Therefore, we define $\rho$ 
to be entangled across $(\alg{R}_{A},\alg{R}_{B})$ just in case 
$\rho|_{\alg{R}_{AB}}$ 
is \emph{not} a weak-$^{*}$ limit of convex combinations of product states of 
$\alg{R}_{AB}$ (Halvorson and Clifton, 2000).      
Notice that we chose weak-$^{*}$ convergence rather than convergence in 
norm, hence we obtain a strong notion of 
entanglement.  In the case $\hil{H}=\hil{H}_{A}\otimes 
\hil{H}_{B}$, $\alg{R}_{A}=\alg{B}(\hil{H}_{A})\otimes I$, and 
$\alg{R}_{B}=I\otimes\alg{B}(\hil{H}_{B})$, the definition obviously 
coincides with the usual notion of entanglement for a pure state 
(convex combinations and approximations being irrelevant in 
that case), and also coincides with
the definition of entanglement (usually called `nonseparability') for a mixed density operator  
that is standard in quantum information theory 
 (Werner, 
1989; Clifton 
and Halvorson, 2000; Clifton \emph{et al}, 2000).  Further evidence 
that the definition captures 
an essentially nonclassical feature of correlations is given by the fact  that  
$\alg{R}_{AB}$ will possess an entangled state in the sense defined 
above
 if and \emph{only if} both $\alg{R}_{A}$ and $\alg{R}_{B}$ are nonabelian 
 (Bacciagaluppi, 1993, Thm. 7; Summers 
 and Werner, 1995, Lemma 2.1). 
Returning to AQFT, it is therefore reasonable to say that
 a global state of the field $\rho$ on $\alg{A}''=\alg{B}(\hil{H})$ is entangled across 
a pair of spacelike-separated regions $(O_{A},O_{B})$ just in case 
$\rho|_{\alg{A}_{AB}}$, $\rho$'s restriction to 
$\alg{A}_{AB}=[\alg{A}(O_{A})\cup\alg{A}(O_{B})]''$, falls outside 
the weak-$^{*}$ closure of the convex hull of $\alg{A}_{AB}$'s product states.

Our next task is to review the mathematical representation of  
operations, highlight some subtleties in their physical 
interpretation, and then discuss what is meant by 
\emph{local} operations on a system.  We then end this section by 
giving the general argument 
that 
local operations performed in either of two spacelike separated 
regions ($O_{A}$,$O_{B}$) cannot 
create entanglement in a state across the regions. 

The most general transformation of the state of a quantum system with 
Hilbert space $\hil{H}$ is described 
by an operation on $\alg{B}(\hil{H})$,  
defined to be a positive, weak-$^{*}$ continuous, linear map $T: 
\alg{B}(\hil{H})\rightarrow\alg{B}(\hil{H})$ satisfying 
$0\leq T(I)\leq I$ (Haag and Kastler, 1964; Davies, 1976; Kraus, 1983; 
Busch \emph{et al}, 1995; Werner 1987).
(The weak-$^{*}$ topology on a von Neumann algebra $\alg{R}$ is 
defined in complete analogy to the weak-$^{*}$ topology on its state space, 
i.e., $\{Z_{n}\}\subseteq \alg{R}$ weak-$^{*}$ converges 
to $Z\in\alg{R}$ just in case 
$\rho(Z_{n})\rightarrow\rho(Z)$ for all 
states $\rho$ of $\alg{R}$.)  
Any such $T$ induces a 
map $\rho\rightarrow\rho^{T}$ from the state space 
of $\alg{B}(\hil{H})$ into itself or $0$, where, for all $Z\in \alg{B}(\hil{H})$,
\begin{equation} 
\rho^{T}(Z)\equiv\rho(T(Z))/\rho(T(I))\ \ \mbox{if}\ 
\rho(T(I))\not=0;\ \ \equiv 0\ \mbox{otherwise}.  
\end{equation}
The number 
$\rho(T(I))$ is the probability that an ensemble 
in state $\rho$ will respond `Yes' to the question represented by the 
positive operator
$T(I)$.  
An operation 
$T$ is called selective if $T(I)<I$, and nonselective if 
$T(I)=I$.  
The final state after a selective operation on an ensemble 
of identically prepared systems  is obtained by ignoring those members of the 
ensemble that fail to respond `Yes' to $T(I)$.  Thus a selective 
operation involves 
performing a physical operation on an ensemble followed by a \emph{purely conceptual}
operation in which one makes a selection of a subensemble based on 
the outcome of the physical operation  (assigning `state' $0$ to the 
remainder).  
Nonselective operations, by contrast, always elicit a `Yes' 
response from any state, hence the final state is not obtained by 
selection but purely as a result of the physical interaction between 
object system and the device that effects the operation.  (We shall 
shortly discuss some actual physical examples to make this general 
description of operations concrete.)  

An operation $T$, which quantum 
information theorists  
call a superoperator (acting, as it does, on operators to 
produce operators), ``can describe any 
combination of unitary operations, interactions with an ancillary 
quantum system or with the environment, quantum measurement, 
classical communication, and subsequent quantum operations 
conditioned on measurement results'' (Bennett \emph{et al}, 1999).  
Interestingly, a superoperator itself can always be represented in terms of 
operators, as 
a consequence of the Kraus representation theorem (1983, p. 42): 
for any operation $T:\alg{B}(\hil{H})\rightarrow\alg{B}(\hil{H})$, 
there exists a (not necessarily unique) countable collection of 
Kraus operators $\{K_{i}\}\subseteq\alg{B}(\hil{H})$ such that
\begin{equation} \label{eq:Kraus}
      T(\cdot)=\sum_{i}K_{i}^{*}(\cdot)K_{i},\ \mbox{with}\ 
      0\leq \sum_{i}K_{i}^{*}K_{i}\leq I
      \end{equation}
      where both sums, if infinite, are to be understood as  
      weak-$^{*}$ convergence.  It is not difficult to show that the sum 
      $\sum_{i}K_{i}K_{i}^{*}$ must also weak-$^{*}$ converge, hence we 
      can let $T^{*}$ denote the operation conjugate to $T$ whose Kraus 
      operators are $\{K_{i}^{*}\}$.  It then follows (using the 
      linearity and cyclicity of the trace) that
       if a state $\rho$ is represented by a density operator $D$ on 
       $\hil{H}$, 
      $\rho^{T}$ will be represented by the density operator $T^{*}(D)$.
      If the mapping $\rho\rightarrow\rho^{T}$, or equivalently, 
      $D\rightarrow T^{*}(D)$, maps pure states to pure states, then 
      the operation $T$ is called a pure operation, and this 
      corresponds to it being representable by a \emph{single} Kraus operator.      
            
            More generally, the Kraus representation shows that a general 
      operation is always equivalent to mixing the results of separating 
      an initial 
      ensemble into subensembles to which one applies pure (possibly 
      selective) operations, represented by the individual Kraus 
      operators.  To see this, let $T$ be an arbitrary operation 
      performed on a 
      state $\rho$, where $\rho^{T}\not=0$, and suppose $T$ is 
      represented by Kraus operators $\{K_{i}\}$.  Let $\rho^{K_{i}}$ 
      denote the result of applying to $\rho$ the pure operation given by the 
      mapping $T_{i}(\cdot)=K_{i}^{*}(\cdot)K_{i}$, and (for 
      convenience) define 
      $\lambda_{i}=\rho(T_{i}(I))/\rho(T(I))$.  Then, at least when 
      there are finitely many Kraus operators, it is easy to see 
      that $T$ itself maps $\rho$ to the convex combination
            $\rho^{T}=\sum_{i}\lambda_{i}\rho^{K_{i}}$.  In the infinite 
            case, this sum converges not just weak-$^{*}$ but \emph{in 
            norm}, and it is a useful exercise in the topologies we 
            have introduced 
            to see why.  Letting $\rho_{n}^{T}$ denote the partial 
            sum $\sum_{i=1}^{n}\lambda_{i}\rho^{K_{i}}$, we need to establish that
            \begin{equation}
             \lim_{n\rightarrow\infty}[\sup\{|\rho^{T}(Z)-\rho_{n}^{T}(Z)|: 
Z=Z^{*}\in\alg{B}(\hil{H}),\ |Z|\leq 1\}]=0.
\end{equation}
For \emph{any} $Z\in\alg{B}(\hil{H})$, we have
\begin{equation}
|\rho^{T}(Z)-\rho_{n}^{T}(Z)|=\rho(T(I))^{-1}|\sum_{i=n+1}^{\infty}\rho(K_{i}^{*}ZK_{i})|.
\end{equation}
However, $\rho(K_{i}^{*}(\cdot)K_{i})$, being a positive linear 
functional, has a norm that may be computed by its action on the 
identity (KR 1997, Thm. 4.3.2).  Therefore, $|\rho(K_{i}^{*}ZK_{i})|\leq 
|Z|\rho(K_{i}^{*}K_{i})$, and we obtain 
\begin{equation}\label{eq:herro}
|\rho^{T}(Z)-\rho_{n}^{T}(Z)|\leq \rho(T(I))^{-1}|Z|\sum_{i=n+1}^{\infty}\rho(K_{i}^{*}K_{i}).
\end{equation}
However, since $\sum_{i}K_{i}^{*}K_{i}$ weak-$^{*}$ converges, this last 
summation is just the tail set of a convergent series.  Therefore, when 
$|Z|\leq 1$, the right-hand side of (\ref{eq:herro}) goes to zero independently of $Z$.  
        
To get a concrete idea of how operations work physically, and to highlight 
two important
interpretational pitfalls, let us again consider our
toy universe, with $\hil{H}=
\mathbb{C}_{A}^{2}\otimes\mathbb{C}_{B}^{2}$ and $x$ an entangled 
state.
Recall that Alice 
disentangled $x$ by measuring an $A$ observable with eigenprojections
$P_{\pm}$.  Her measurement corresponds to applying the nonselective 
operation $T$ 
with Kraus operators $K_{1}=P_{+}\otimes I$ and $K_{2}=P_{-}\otimes 
I$, resulting in the final state $T^{*}(P_{x})= T(P_{x})=D'$, as given in 
(\ref{eq:Alice1}).  If Alice were to further 
`apply' the pure 
selective operation $T'$ represented by the single Kraus operator $P_{+}\otimes 
I$, the final state of her 
ensemble, as is apparent from (\ref{eq:Alice2}), would be the product state 
$D''=P_{+}\otimes 
P_{b_{x}^{+}}$.  But, as we have emphasized, this corresponds 
to a conceptual operation in which Alice just throws away all members of 
the original ensemble that yielded measurement outcome $-1$.  
On the 
other hand, it is essential not lose sight of the issue that troubled Einstein. 
\emph{Whatever} outcome Alice selects for, she will then be in a 
position to assert that certain $B$ observables --- those that have either 
$b_{x}^{+}$ or $b_{x}^{-}$ as an eigenvector, depending on the outcome 
she favours --- have a 
sharp value in the ensemble she is left with. 
But  prior to Alice performing the first operation $T$, such an  
assertion would have contradicted the orthodox interpretation of the 
entangled superposition $x$.  If, contra Bohr, one were to view this change in 
$B$'s state as a \emph{real physical} change brought about by one of the operations Alice performs, 
surely the innocuous conceptual operation $T'$ could not be the 
culprit --- it must have been $T$ which forced
$B$ to `choose' between the alternatives $b_{x}^{\pm}$.  
Unfortunately, this clear distinction between the physical operation $T$ and conceptual 
operation $T'$ is not reflected well in the formalism
of operations.  For we could equally well have represented Alice's final product state 
$D''=P_{+}\otimes 
P_{b_{x}^{+}}$, not as the result of successively applying the 
operations $T$ and 
$T'$, but as the outcome of applying the single composite operation 
$T'\circ T$, which is just the mapping $T'$.  And \emph{this} $T'$ now needs to be 
understood, not purely as a conceptual operation, but as also involving a physical operation, 
with possibly real nonlocal effects on $B$, depending on one's view of the EPR 
paradox.  (In particular, keep in mind that you are taking the first step on the road to 
conceding the incompleteness of quantum theory if you attribute 
the change in the state of $B$ brought about 
by $T'$ in this case to a mere change in 
Alice's \emph{knowledge} about $B$'s state.) 

There is a second pitfall that 
concerns interpreting the result of 
\emph{mixing} 
subensembles, as opposed to singling out a particular 
subensemble. Consider an alternative method available to 
Alice for disentangling a state $x$.  For 
concreteness, let us suppose that $x$ is the singlet state 
$1/\sqrt{2}(a^{+}\otimes b^{-}-a^{-}\otimes b^{+})$. Alice applies the 
nonselective operation with Kraus representation
\begin{equation} \label{eq:Alice4}
T(\cdot)=\frac{1}{2}(\mathbf{\sigma}_{a}\otimes I) (\cdot 
)(\mathbf{\sigma}_{a}\otimes 
I)+\frac{1}{2}(I\otimes I) (\cdot)(I\otimes I),
\end{equation} 
where $\mathbf{\sigma}_{a}$ is the spin observable with eigenstates $a^{\pm}$.
 Since $\mathbf{\sigma}_{a}\otimes I$ maps $x$ to the triplet state 
 $1/\sqrt{2}(a^{+}\otimes b^{-}+a^{-}\otimes b^{+})$, $T^{*}$ ($=T$) 
 will map $P_{x}$ 
to an equal mixture of the singlet and triplet, which admits the 
following convex decomposition into product states
\begin{equation}  \label{eq:Alice3}
D' = \frac{1}{2}P_{a^{+}\otimes b^{-}}+\frac{1}{2}P_{a^{-}\otimes 
b^{+}}.
\end{equation}
Has Alice truly disentangled $A$ from $B$?  Technically, Yes.
Yet all Alice has done, physically, is to separate the initial $A$ ensemble into 
two subensembles in equal proportion, left the second 
subensemble alone while performing a (pure, nonselective) unitary operation 
$\mathbf{\sigma}_{a}\otimes I$ 
on the 
first that maps all its $A+B$ pairs to the triplet state, and then 
remixed the ensembles.  Thus, notwithstanding the above decomposition of the 
final density matrix $D'$, Alice \emph{knows quite well} that she is 
in possession of an ensemble of $A$ systems each of which is 
entangled either via the singlet or triplet state with the 
corresponding $B$ systems.  This will of course be recognized as one 
aspect of the problem with the ignorance interpretation of mixtures.  
We have two different ways to decompose $D'$ --- as an equal mixture 
of the singlet and triplet or of two product states --- but which is 
the correct way to understand how the ensemble is \emph{actually} 
constituted?  The definition of entanglement is just not sensitive to the 
answer.  (It is exactly this insensitivity that is at 
the heart of the recent dispute over whether NMR quantum computing 
is correctly understood as implementing genuine \emph{quantum} computing that cannot be simulated 
classically (Braunstein \emph{et al}, 1999; Laflamme, 1998).)  Nevertheless, we are inclined to think the destruction of 
the singlet's 
entanglement that 
Alice achieves by applying the 
operation in (\ref{eq:Alice4}) is an artifact of her mixing process, in which she is 
represented as simply forgetting 
about the history of the $A$ systems.  
And this is the view we shall take when we consider similar 
possibilities for destroying entanglement between field systems in AQFT. 
  
In the two examples considered above, 
Alice
applies operations whose Kraus operators 
lie in the subalgebra $\alg{B}(\hil{H}_{A})\otimes I$ 
associated with system $A$.  
In the case of a nonselective operation, this is clearly sufficient
for her operation not to have any effect on the 
expectations of the observables of system $B$.  However, it is also necessary.  The point is quite 
general. Let us define a 
nonselective operation $T$ to be (\emph{pace} Einstein!) local to the subsystem
represented by a von Neumann subalgebra 
$\alg{R}\subseteq\alg{B}(\hil{H})$ just in case  
$\rho^{T}|_{\alg{R'}}=\rho|_{\alg{R'}}$ for all states $\rho$.  Thus, 
we require that $T$ 
leave the expectations of observables outside of $\alg{R}$, as well 
as those in its 
center $\alg{R}\cap\alg{R}'$, unchanged.  Since distinct 
states of $\alg{R}'$ cannot 
agree on all expectation values, this means $T$ must act like 
the identity operation on $\alg{R}'$.  Now fix an arbitrary 
element $Y\in\alg{R}'$, and 
suppose $T$ is represented by Kraus operators $\{K_{i}\}$.  A 
straightforward calculation reveals that
\begin{equation} \label{eq:werner}
\sum_{i}[Y,K_{i}]^{*}[Y,K_{i}]=T(Y^{2})-T(Y)Y-YT(Y)+YT(I)Y.
\end{equation}
Since $T(I)=I$, and $T$ leaves the elements of $\alg{R}'$ fixed, 
the right-hand side of (\ref{eq:werner}) reduces to zero.  
Thus each of the terms in the sum on 
the left-hand side, which are positive operators, must 
individually be zero.  Since $Y$ was an arbitrary element 
of $\alg{R}'$, it 
follows that $\{K_{i}\}\subseteq (\alg{R}')'=\alg{R}$.  So we see that 
nonselective operations local to $\alg{R}$ \emph{must} be represented by Kraus 
operators taken from the subalgebra $\alg{R}$.  

As for selective 
operations, we have already seen that they
\emph{can}
`change' the global statistics of  a state $\rho$ outside the 
subalgebra $\alg{R}$, particularly when $\rho$ is entangled.  
However, a natural extension of the definition of local operation on 
$\alg{R}$ to a cover the case when $T$ is selective is to require that 
$T(Y)=T(I)Y$ for all $Y\in\alg{R}'$.  This implies 
$\rho^{T}(Y)=\rho(T(I)Y)/\rho(T(I))$, and so guarantees that $T$ will 
leave the statistics of any observable in $\alg{R}'$ the same 
\emph{modulo} whatever correlations that observable might have had in the initial 
state with the Yes/No 
question represented by the 
positive operator $T(I)$.  Further motivation is provided by the fact 
this definition is equivalent to requiring that $T$ factor across the 
algebras $(\alg{R},\alg{R}')$, in the sense that $T(XY)=T(X)Y$ for 
all $X\in\alg{R}$, $Y\in\alg{R}'$ (Werner, 1987, Lemma).  If there exist product states across 
$(\alg{R},\alg{R}')$ (an assumption we shall later see does \emph{not} 
usually hold when $\alg{R}$ is a local algebra in AQFT), this guarantees that any local selective 
operation on $\alg{R}$, when the global state is an entirely 
uncorrelated product state, will leave the statistics of that state on 
$\alg{R}'$ unchanged.  Finally, observe that 
$T(Y)=T(I)Y$ for all $Y\in\alg{R}'$ implies that the right-hand side of (\ref{eq:werner}) again 
reduces to zero.  Thus it follows (as before) that  
selective local operations 
on $\alg{R}$ must also be represented by Kraus 
operators taken from the subalgebra $\alg{R}$.      

Applying these considerations to field theory, any local operation on the field 
system 
within a region $O$, whether or not the operation is selective, is represented by a family of Kraus 
operators taken from $\alg{A}(O)$.  In particular, 
each individual
element of $\alg{A}(O)$ represents a pure operation that can be 
performed within $O$ (cf. Haag and Kastler, 1964, p. 850).    
We now need to argue that local operations performed by two 
experimenters in spacelike separated regions cannot create entanglement 
in a state across the regions where it had none before.  This point, well-known 
 by quantum information theorists working in nonrelativistic quantum theory, 
 in fact applies 
quite generally to any two commuting von Neumann algebras $\alg{R}_{A}$ and 
$\alg{R}_{B}$.  

Suppose that a state  
$\rho$ is not entangled across $(\alg{R}_{A},\alg{R}_{B})$, local 
operations $T_{A}$ and $T_{B}$ are applied to $\rho$, and the result is 
nonzero (i.e., some members of the initial ensemble are not discarded).  
Since the Kraus 
operators of these operations commute, it is easy to check that 
$(\rho^{T_{A}})^{T_{B}}=(\rho^{T_{B}})^{T_{A}}$, so it does not 
matter in 
which order we take the operations.  It is sufficient to show 
that $\rho^{T_{A}}$ will again be unentangled, for then we can just 
repeat the same argument to obtain that neither can $(\rho^{T_{A}})^{T_{B}}$ 
be entangled.  Next, recall that a general operation $T_{A}$ will just produce a 
mixture over the results of applying a countable collection of pure operations 
to $\rho$; more precisely, the result will be the norm, and hence weak-$^{*}$, limit of 
finite convex combinations of the results of applying pure operations 
to $\rho$.  
If the states that result from $\rho$ under those pure operations 
are themselves not 
entangled, $\rho^{T_{A}}$ itself could not be either, because the set of 
unentangled states is by definition convex and weak-$^{*}$ closed.  Without loss of 
generality, then, we may assume that the local operation $T_{A}$ is pure and, hence, given by 
$T_{A}(\cdot)=K^{*}(\cdot)K$, for some
\emph{single} Kraus operator 
$K\in\alg{R}_{A}$.  As before, we shall denote the resulting 
state $\rho^{T_{A}}$ by $\rho^{K}$ ($\equiv \rho(K^{*}\cdot 
K)/\rho(K^{*}K)$).  

Next, suppose that $\omega$ is any product state on 
$\alg{R}_{AB}$ with restrictions to $\alg{R}_{A}$ and 
$\alg{R}_{B}$ given by $\omega_{A}$ and $\omega_{B}$, and such that 
$\omega^{K}\not=0$.  Then, for any 
$X\in\alg{R}_{A}$, $Y\in\alg{R}_{B}$,
\begin{eqnarray}
 \omega^{K}(XY) & = & 
 \frac{\omega(K^{*}(XY)K)}{\omega(K^{*}K)} \\
 & = & \frac{\omega(K^{*}XKY)}{\omega(K^{*}K)} \\ 
 & = & \frac{\omega_{A}(K^{*}XK)}{\omega_{A}(K^{*}K)}\omega_{B}(Y)
 = \omega_{A}^{K}(X)\omega_{B}(Y).
 \end{eqnarray}
It follows that $K$ maps product states of $\alg{R}_{AB}$ to 
product states (or to zero).
Suppose, instead, that $\omega $ is a convex combination of states on 
$\alg{R}_{AB}$,
$\omega =\sum _{i=1}^{n}\lambda _{i}\omega _{i}$.  
Then, setting
$\lambda ^{K}_{i}= \omega _{i}(K^{*}K)/\omega (K^{*}K)$, it is 
easy to see that
$\omega ^{K}=\sum _{i=1}^{n}\lambda ^{K}_{i}\omega _{i}^{K}$,
hence $K$ preserves convex combinations of states on 
$\alg{R}_{AB}$ as well.  It is also not difficult to see that the mapping 
$\omega\mapsto\omega^{K}$ is weak-$^{*}$ continuous at any point where 
$\omega^{K}\not=0$  
 (cf. Halvorson and Clifton, 2000, Sec. 3).  
Returning to our original state $\rho$, our hypothesis is that it is 
 not 
entangled. Thus, there is a net 
of states 
$\{\omega_{n}\}$ on $\alg{R}_{AB}$, each of which is a convex combination of product states, 
such that 
$\omega_{n}\rightarrow \rho|_{\alg{R}_{AB}}$ in the weak-$^{*}$ 
topology.  It follows from the 
above considerations that 
$\omega_{n}^{K}\rightarrow \rho^{K}|_{\alg{R}_{AB}}$, where 
each of the states $\{\omega_{n}\}$ is again a convex combination of 
product states.  Hence $\rho^{K}|_{\alg{R}_{AB}}$ is not 
entangled either.

     \section*{3. The Operational Implications of the \\ Reeh-Schlieder 
      Theorem}   
       
     Again, let $\alg{R}\subseteq\alg{B}(\hil{H})$ be any von Neumann 
     algebra.  A vector $x\in\hil{H}$ is called cyclic for $\alg{R}$ 
     if the norm closure of the set $\{Ax:A\in\alg{R}\}$
          is the \emph{whole} of $\hil{H}$. In AQFT, the Reeh-Schlieder 
     (RS) theorem 
     connects this formal property of cyclicity to the physical 
     property of a field state having bounded energy.  (More 
     generally, the connection is between cyclicity and field states that 
     are `analytic' in the energy.  This, together with the physical and 
     mathematical origins of the RS theorem, are analyzed in depth in Halvorson (2000).)  A pure global state $x$ of the field 
     has bounded energy just in case $E([0,r])x=x$ for some 
     $r<\infty$, where $E$ is the spectral measure for the global 
     Hamiltonian of the field.  In other words, the probability in 
     state $x$  
     that the field's energy is confined to the bounded interval 
     $[0,r]$ is unity.  In particular, the vacuum $\Omega$ is an eigenstate 
     of the Hamiltonian 
     with eigenvalue $0$, and hence trivially has bounded energy.  The 
     RS theorem implies that 
     \begin{center}
     \emph{If $x$ has bounded energy, then x is cyclic for any local
        algebra $\alg{A}(O)$.}\end{center}
       \noindent Our first order of business is to explain Streater and Wightman's 
 comment that the RS 
     theorem entails ``it is difficult 
to isolate a system described by fields from outside effects'' (1989, p. 
139).  

A vector $x$ is called separating for a von Neumann algebra $\alg{R}$ if $Ax=0$ 
implies $A=0$ whenever $A\in\alg{R}$.  It is an elementary result
of von Neumann algebra theory that $x$ will be cyclic for 
$\alg{R}$ if and only if $x$ is separating for $\alg{R}'$ (KR 1997, Cor. 
5.5.12).   To illustrate with a simple example, take 
$\hil{H}=\hil{H}_{A}\otimes \hil{H}_{B}$.  If $\dim\hil{H}_{A}\geq\dim\hil{H}_{B}$, 
then it is possible for there to be vectors $x\in\hil{H}$ that have a Schmidt 
decomposition $\sum_{i}c_{i}a_{i}\otimes b_{i}$ 
where $|c_{i}|^{2}\not=0$ for \emph{all} $i=1$ to $\dim\hil{H}_{B}$.  If we act on such an $x$ by an 
operator in the subalgebra $I\otimes\alg{B}(\hil{H}_{B})$, of form 
$I\otimes B$, then the only way $(I\otimes B)x$ can be the zero vector is if 
$B$ itself maps all the basis vectors $\{b_{i}\}$ to zero, i.e., $I\otimes 
B=0$.  Thus such vectors are separating for 
$I\otimes\alg{B}(\hil{H}_{B})$, and therefore cyclic for 
$\alg{B}(\hil{H}_{A})\otimes I$.  Conversely, it is easy to convince 
oneself that 
$\alg{B}(\hil{H}_{A})\otimes I$ possesses a cyclic vector 
--- equivalently, $I\otimes\alg{B}(\hil{H}_{B})$  has a separating 
vector --- 
\emph{only if} $\dim\hil{H}_{A}\geq\dim 
\hil{H}_{B}$.  So, to take another example, each of the $A$ and $B$ subalgebras 
will possess a cyclic and a separating vector just in case  
$\hil{H}_{A}$ and 
$\hil{H}_{B}$ have the same dimension (cf. the proof of Clifton \emph{et 
al} 1998, Thm. 4).  

Consider, now, a local algebra 
$\alg{A}(O)$ with $O'\not=\emptyset$, and a field state $x$ with 
bounded energy.  
The RS theorem tells us that $x$ is cyclic for 
$\alg{A}(O')$, and therefore, separating for $\alg{A}(O')'$.  But by 
microcausality, $\alg{A}(O)\subseteq\alg{A}(O')'$, hence $x$ must be 
separating for the subalgebra $\alg{A}(O)$ as well.  Thus it is an 
immediate corollary to the 
RS theorem that 
\begin{center}\emph{If $x$ has bounded energy, then x is separating for any local
        algebra $\alg{A}(O)$ with $O'\not=\emptyset$.}\end{center}
 \noindent It is this corollary that prompted Streater and Wightman's remark.  
But what has it got to do with thinking of the field system 
$\alg{A}(O)$ as isolated?
For a start, we can now show that the local restriction $\rho_{x}|_{\alg{A}(O)}$ 
of a state with bounded energy is always a highly `noisy' mixed 
state.  Recall that a state $\omega$ on a von Neumann algebra 
$\alg{R}$ is a component of another state $\rho$ if there is a third 
state $\tau$ such that $\rho=\lambda\omega+(1-\lambda)\tau$ with 
$\lambda\in (0,1)$ (Van Fraassen 1991, p. 161).  We are going 
to show that $\rho_{x}|_{\alg{A}(O)}$ has a \emph{norm} dense set of 
components in the state space of $\alg{A}(O)$.  

Once again, the point is quite general.  Let $\alg{R}$ be any von 
Neumann algebra, $x$ be separating for $\alg{R}$, and let $\omega$ be 
an arbitrary state of $\alg{R}$.  We must find a
sequence
$\{ \omega _{n} \}$ of states of $\alg{R}$ such that each $\omega _{n}$ is a
component of
$\rho_{x}|_{\alg{R}}$ and $\|\omega _{n}-\omega \|\rightarrow 0$.
Since $\alg{R}$ has a separating vector, it follows that every state of
$\alg{R}$ is a vector state (KR 1997, Thm 7.2.3).  (That this should 
be so is not 
as surprising as it sounds.  Again, if 
$\hil{H}=\hil{H}_{A}\otimes \hil{H}_{B}$, and $\dim\hil{H}_{A}\geq\dim 
\hil{H}_{B}$, then as we have seen, the B subalgebra possesses a 
separating vector.  But it is also easy to see, in this case, that every state on 
$I\otimes\alg{B}(\hil{H}_{B})$ is the reduced density operator 
obtained from a pure state on 
$\alg{B}(\hil{H})$ determined by a vector in $\hil{H}$.)   
In particular, there is a nonzero vector $y\in \hil{H}$
such that
$\omega =\omega _{y}$.  Since $x$ is separating for $\alg{R}$, $x$ is cyclic for
$\alg{R}'$, therefore we may choose a sequence of
operators $\{A_{n} \} \subseteq \alg{R}'$ so that $A_{n}x\rightarrow y$.  
Since $|A_{n}x-y|\rightarrow 0$, $\|\omega
    _{A_{n}x}-\omega _{y}\|\rightarrow 0$.  We claim now that each
  $\omega _{A_{n}x}$ is a component of $\rho_{x}|_{\alg{R}}$.  Indeed, for any
  positive element $B^{*}B\in\alg{R}$, we have:
  \begin{eqnarray}
    \langle A_{n}x,B^{*}BA_{n}x\rangle &=&\langle
    x,A_{n}^{*}A_{n}B^{*}Bx\rangle
    \:=\: \langle Bx,A_{n}^{*}A_{n}Bx\rangle \\
    &\leq &|A_{n}^{*}A_{n}|\langle Bx,Bx\rangle \: =\:
    |A_{n}|^{2}\langle x,B^{*}Bx\rangle . \end{eqnarray} Thus, 
    \begin{eqnarray} \label{eq:positive}
    \omega _{A_{n}x}(B^{*}B) &=& \frac{\langle
      A_{n}x,B^{*}BA_{n}x\rangle }{|A_{n}x|^{2}}\:\leq
    \:\frac{|A_{n}|^{2}}{|A_{n}x|^{2}}\,\rho _{x}(B^{*}B)
    .\end{eqnarray} If we now take $\lambda =
  |A_{n}x|^{2}/|A_{n}|^{2}\in (0,1)$, and consider 
  the linear functional  
  $\tau$ on $\alg{R}$ given by 
  $\tau=(1-\lambda)^{-1}(\rho_{x}|_{\alg{R}}-\lambda\omega_{A_{n}x})$, 
   then (\ref{eq:positive}) 
  implies that $\tau$ is a 
  state (in particular, positive), and we see that $\rho _{x}|_{\alg{R}}=\lambda\omega
  _{A_{n}x}+(1-\lambda)\tau$ as required. (This result also holds more 
  generally for 
  states $\rho$ of $\alg{R}$ that are faithful, i.e., 
  $\rho(Z)=0$ entails $Z=0$ for any positive $Z\in\alg{R}$;
  see the first part of the 
  proof of Summers and Werner, 1988, Thm. 
  2.1.)
  
  So bounded energy states are, locally, highly mixed.  And such 
  states are far from special --- they lie norm dense in the pure 
  state space of  $\alg{B}(\hil{H})$.  To see this, just recall that it is part of the 
  spectral theorem for the global Hamiltonian that 
  $E([0,n])$ converges strongly to the identity as 
  $n\rightarrow\infty$.  Thus we may 
  approximate any vector $y\in\hil{H}$ by the sequence of bounded 
  energy states
  $\{E([0,n])y/|E([0,n])y|\}_{n=0}^{\infty}$.  Since there are so many bounded 
  energy states of the field, that are locally so `noisy',  Streater and 
  Wightman's comment is entirely warranted.
  But somewhat more can be said.  As we saw with our toy 
  example in Section 1, when a local subsystem of a global system in a pure 
  state is itself in a mixed state, this is a sign 
  of that subsystem's entanglement with its environment. And there is 
  entanglement lurking in bounded energy states too.  
  But, first, 
  we need to take a closer look at the operational implications of 
  local cyclicity. 
  
  If a vector $x$ is cyclic for 
  $\alg{R}$, then for any $y\in\alg{H}$, 
  there is a sequence $A_{n}\in\alg{R}$ such that $A_{n}x\rightarrow 
  y$.  Thus for any  
  $\epsilon>0$ there is an $A\in\alg{R}$ such that 
  $\|\rho_{Ax}-\rho_{y}\|<\epsilon$.  However, $\rho_{Ax}$ 
  is just 
  the state one gets by applying the pure 
  operation given by the Kraus operator $K=A/|A|\in\alg{R}$ to $\rho_{x}$.  It follows 
  that if $x$ is cyclic for $\alg{R}$, one can get arbitrarily close 
  in norm to any other pure state of $\alg{B}(H)$ by applying an 
  appropriate pure local operation in $\alg{R}$ to $\rho_{x}$.  In particular, pure 
  operations on the vacuum $\Omega$ within a local region $O$, no matter how small, can prepare 
  essentially any global state 
  of the field.  As Haag emphasizes, to do this the 
  operation must 
  ``judiciously exploit the small but nonvanishing long 
distance correlations which exist in the vacuum'' (1992, p. 102).  
This, as Redhead (1995) has argued by analogy to the 
singlet state, is made possible by the fact 
that the vacuum is highly entangled (cf. Clifton \emph{et 
al} 1998).  But the first puzzle we need 
to sort out is that it looks as though entirely \emph{physical} operations 
in $O$ can change the global state, in particular the vacuum $\Omega$, 
to any desired state!  (For example, Segal and 
Goodman (1965) have called this  
``bizarre'' and ``physically quite 
surprising'', sentiments echoed recently by
Fleming who calls it ``amazing!'' (1999).)

Redhead's analysis of the cyclicity of 
the singlet state $1/\sqrt{2}(a^{+}\otimes b^{-}-a^{-}\otimes b^{+})$ for 
the subalgebra $\alg{B}({\mathbb{C}}_{A}^{2})\otimes I$
is designed to remove this puzzle (\emph{ibid}, p. 128).  
(Note that in this simple $2\times 2$-dimensional case, he could equally well have 
chosen \emph{any} entangled state, since they are all separating for 
$I\otimes\alg{B}({\mathbb{C}}_{B}^{2})$.)  Redhead writes:
\begin{quote}
``\ldots we 
 want to distinguish clearly two senses of the term ``operation''.  
 Firstly there are physical operations such as making measurements, 
 selecting subensembles according to the outcome of measurements, and 
 mixing ensembles with probabilistic weights, and secondly there are 
 the mathematical operations of producing superpositions of states by 
 taking linear combinations of pure states produced by appropriate 
 selective measurement procedures.  These superpositions are of course 
 quite different from the mixed states whose preparation we have 
 listed as a physical operation'' (1995, pp. 128-9).
\end{quote}
Note that, in stark contrast to our discussion in the previous section,
 Redhead counts selecting subensembles and mixing as 
physical operations; it is only the operation of 
superposition that warrants the adjective `mathematical'.  When he 
explains why it is possible that $x$ can be cyclic, Redhead first 
notes (\emph{ibid}, p. 129) that the four basis states 
\begin{equation} \label{eq:states}
a^{+}\otimes b^{-},\ a^{-}\otimes b^{-},\ a^{-}\otimes b^{+},\ a^{+}\otimes 
b^{-},
\end{equation}
are easily obtained by the physical operations of applying 
projections and unitary transformations to the singlet state, and 
exploiting the fact that the singlet strictly correlates $\sigma_{a}$ with $\sigma_{b}$.  He goes on:
\begin{quote}      
``But \emph{any} state for the joint system is some linear combination 
of these four states, so by 
 the \emph{mathematical} operation of linear combination, we can see 
 how to generate an arbitrary state in 
 $\hil{H}_{1}\otimes\hil{H}_{2}$ from physical operations performed 
 on particle one.  But all the operations we have described can be 
 represented in the algebra of operators on $\hil{H}_{1}$ (extended to 
 $\hil{H}_{1}\otimes\hil{H}_{2}$)''  (\emph{ibid}, p. 129).
 \end{quote}
 
 Now, while Redhead's explanation of why it is mathematically possible 
 for $x$ to be cyclic is perfectly correct, he actually misses the 
 mark when it comes to the physical interpretation of cyclicity.  The 
 point is
 that superposition \emph{of states} is a 
 red-herring.  Certainly a superposition 
 of the states in (\ref{eq:states}) could not be prepared by physical operations confined to the $A$ 
 system.  But, as Redhead himself notes in the final 
 sentence above, one can get the same \emph{effect} as superposing those 
 states by acting on $x$ with an operator of form $A\otimes I$ in the subalgebra 
 $\alg{B}({\mathbb{C}}_{A}^{2})\otimes I$ --- an operator that is itself a `superposition' of other operators 
 in that algebra.  
 What Redhead fails to point out
  is that the action of this operator on $x$ \emph{does have a 
  local physical
 interpretation}: as we have seen, it is a Kraus operator that represents the outcome of a generalized 
 positive operator valued measurement on the $A$ system.  The key to 
 the puzzle is, rather, that this positive operator valued measurement will generally have 
 to be 
 \emph{selective}.  For one certainly could never, with nonselective 
 operations on $A$ alone, get as close as one likes to any state vector in 
 $\mathbb{C}_{A}^{2}\otimes\mathbb{C}_{B}^{2}$ (otherwise all state 
 vectors would induce the same state on 
 $I\otimes\alg{B}({\mathbb{C}}_{B}^{2})$!). We conclude that the 
 correct way to 
 view the physical content of cyclicity is that changes in the global 
 state are partly due to an experimenter's ability to perform a generalized measurement on 
 $A$, and partly due (\emph{pace} Redhead) to the purely 
 conceptual operation of selecting a subensemble based on the outcome 
 of the experimenter's measurement together with the consequent `change' in the state 
 of $B$ via the EPR correlations between $A$ and $B$.
 
One encounters the same interpretational pitfall concerning the cyclicity of the vacuum 
in relation to localized states in AQFT.  A global state of the 
field is said to be localized in $O$ if its expectations on the 
algebra $\alg{A}(O')$ agree with vacuum expectation values (Haag, 1992, 
p. 102).  Thus localized states are `excitations' of the vacuum 
confined to $O$.  In particular, $U\Omega$ will be a localized 
state whenever $U$ is a unitary operator taken from  
$\alg{A}(O)$ (since unitary operations are nonselective). 
But every element of a $C^{*}$-algebra is a finite linear combination of 
unitary operators (KR 1997, Thm. 4.1.7).  
Since $\Omega$ is cyclic for $\alg{A}(O)$, this means we must be able 
to approximate any global state by linear superpositions of vectors describing states 
localized in $O$ --- even approximate states that are localized in regions spacelike 
separated with $O$! Haag, 
rightly cautious, calls this 
a ``(superficial) paradox'' (1992, p. 254), but he fails to put his finger on its 
resolution: while unitary operations are 
nonselective, a local operation in $\alg{A}(O)$ given by a Kraus 
operator that is a linear combination of 
local unitary operators will generally be \emph{selective}. (Haag \emph{does} 
make the interesting point out that only a proper 
subset of 
the state space of a field can be approximated if we restrict 
ourselves to local operations that involve a physically reasonable expenditure 
of energy.  But we do not share the view of Schroer (1999) that
 this point by itself reconciles the RS theorem with 
`common sense'.)

The (common) point of the previous two paragraphs is perhaps best 
summarized as follows.  
Both Redhead and Haag would agree that unitary Kraus operators in 
$\alg{A}(O)$ give rise to purely physical 
operations in the local region $O$.  But there are many Kraus operators in $\alg{A}(O)$ that 
do not represent purely physical operations in $O$ insofar as they are 
selective.  Since every Kraus operator is a linear superposition of 
unitary operators, it follows that ``superposition of local operations'' 
does not preserve (pure) physicality.  Redhead is right that the key to 
diffusing the paradox is in noting that superpositions are 
involved --- but it is essential to understand these superpositions as 
occurring locally in 
$\alg{A}(O)$, not in the Hilbert space.     

Our next order of business is to supply the rigorous argument behind Redhead's intuition about 
the connection between cyclicity and entanglement.  The point, again, 
is quite general: for any two commuting nonabelian von Neumann algebras $\alg{R}_{A}$ and 
$\alg{R}_{B}$, and any state vector $x$ cyclic for $\alg{R}_{A}$ 
(or $\alg{R}_{B}$), $\rho_{x}$ will be entangled across the algebras 
(Halvorson and Clifton, 2000, Prop. 2). For suppose, in order to 
extract a contradiction, that $\rho_{x}$ is \emph{not} 
entangled.  Then as we have seen, operations on $\rho_{x}$ that are local to $\alg{R}_{A}$ cannot 
turn that state into an entangled state across 
$(\alg{R}_{A},\alg{R}_{B})$.  Yet, by the cyclicity of $x$, we know that we can apply pure operations 
to $\rho_{x}$, that are local to 
$\alg{R}_{A}$ 
(or $\alg{R}_{B}$),  and approximate in norm (and hence weak-$^{*}$ 
approximate) any other vector state of 
$\alg{R}_{AB}$.  It follows that no vector state of $\alg{R}_{AB}$ 
could be entangled across $(\alg{R}_{A},\alg{R}_{B})$, and the same goes for 
all its mixed states, which lie in the the norm closed convex hull of the 
vector states.  
But this means that 
$\alg{R}_{AB}$ would possess \emph{no} entangled states at all --- in 
flat 
contradiction with the fact that neither $\alg{R}_{A}$ nor 
$\alg{R}_{B}$ is abelian.  

Returning to the context of AQFT, if we now consider \emph{any} two spacelike 
separated field systems, $\alg{A}(O_{A})$ and $\alg{A}(O_{B})$, then 
the argument we just gave establishes that the dense set of field 
states bounded in the energy will \emph{all} be entangled across the regions 
$(O_{A},O_{B})$.  (Note that the fact that $\alg{A}(O_{A})$ and 
$\alg{A}(O_{B})$
are nonabelian is \emph{itself} a consequence of the RS theorem.  For 
if, say, 
$\alg{A}(O_{A})$ were abelian, then since by the RS theorem that algebra  possesses a cyclic 
vector, it must be a maximal abelian subalgebra of $\alg{B}(\hil{H})$ (KR 1997, 
Cor. 7.2.16).  The same conclusion would have to follow for any 
subregion $\tilde{O}_{A}\subset O_{A}$ whose closure is a proper 
subset of $O_{A}$.  
And this, by isotony, 
would lead to the absurd 
conclusion that $\alg{A}(\tilde{O}_{A})=\alg{A}(O_{A})$, which is 
easily shown to be inconsistent with the axioms of AQFT (Horuzy, 1988, 
Lemma 1.3.10).)  
However, by itself this result does not 
imply that Alice cannot destroy a bounded energy state $x$'s entanglement 
across $(O_{A}, O_{B})$ by performing local operations in $O_{A}$.  
In fact, Borchers (1965, Cor. 7) has shown
that any vector state of form $Ax$ for any nontrivial $A\in \alg{A}(O_{A})$ \emph{never} 
has bounded energy (nor is `analytic' in the energy).  So it seems that all 
Alice needs to do is perform any pure operation within $O_{A}$ and the 
resulting state, because it is no longer subject to the RS theorem, 
need no longer be entangled across $(O_{A}, O_{B})$.  

However, the RS theorem gives only a sufficient, \emph{not} a 
necessary, condition for a 
state $x$ of the field to be cyclic for $\alg{A}(O_{A})$.   And 
notwithstanding that no pure operation Alice performs can preserve boundedness in 
the energy, \emph{almost all} the pure operations she could perform \emph{will} preserve the 
state's cyclicity!  The reason is, 
once again, quite general. Again let $\alg{R}_{A}$ and 
$\alg{R}_{B}$ be two commuting nonabelian von Neumann algebras,  
suppose $x$ is cyclic for $\alg{R}_{A}$, and consider the state 
induced by the vector 
$Ax$ 
where $A\in \alg{R}_{A}$.  Now every element in a von Neumann algebra is 
the strong limit of invertible elements in the algebra (Dixmier and 
Mar\'{e}chal, 1971, Prop. 1).  Therefore, there is a sequence of 
invertible operators $\{\tilde{A}_{n}\}\subseteq \alg{R}_{A}$ such that 
$\tilde{A}_{n}x\rightarrow Ax$, i.e., $\|\rho_{\tilde{A}_{n}x}-
\rho_{Ax}\|\rightarrow 0$.  Notice, however, that since each $\tilde{A}_{n}$ is 
invertible, each vector $\tilde{A}_{n}x$ is again cyclic for 
$\alg{R}_{A}$, 
because we can `cycle back' to $x$ by applying to 
$\tilde{A}_{n}x$ the 
inverse operator $\tilde{A}_{n}^{-1}\in\alg{R}_{A}$, and 
from there we know, by hypothesis, that we can cycle with elements of $\alg{R}_{A}$ arbitrarily 
close to any other vector in 
$\hil{H}$.   It follows that, even 
though 
Alice may \emph{think} she has applied the pure operation given by 
some Kraus operator $A/|A|$ 
to $x$, she 
could well have \emph{actually} applied an invertible Kraus operation given by one of 
the operators $\tilde{A}_{n}/|\tilde{A}_{n}|$ 
in a strong neighborhood of $A/|A|$.  And if she actually did this, then she 
certainly would \emph{not} disentangle $x$, because she would not have 
succeeded in destroying the \emph{cyclicity} of the field state for her local algebra.
We could, of course, give Alice the 
freedom to employ more general mixing operations in $O_{A}$.  But as we saw in the last 
section, it is far from
clear whether a mixing operation should count as a successful 
disentanglement when all the states that are mixed by her operation are 
themselves 
entangled --- or at least not \emph{known} by Alice to be disentangled (given her 
practical inability to specify exactly which Kraus operations go into
the pure operations of her mixing process).  

Besides this, there is a 
more fundamental practical limitation facing Alice, even if we allow 
her any local operation she chooses.  If, as we have seen, we can approximate  
the result of acting on $x$ with any 
given operator in von Neumann algebra $\alg{R}$ by acting on $x$ with an invertible operator that 
preserves $x$'s cyclicity, then  
the set of all such `invertible actions' on $x$ must itself produce a 
dense set of vector states, given that $\{Ax:A\in\alg{R}\}$ is dense.  It follows that if a von 
Neumann algebra possesses even just one cyclic vector, it must possess a 
dense set of them (Dixmier and 
Mar\'{e}chal, 1971, Lemma 4; cf. Clifton \emph{et 
al} 1998).  Now consider, again,
the general situation of two commuting nonabelian algebras $\alg{R}_{A}$ and 
$\alg{R}_{B}$, where either algebra possesses a cyclic vector, and hence a 
dense set of such.  If, in addition, the 
algebra $\alg{R}_{AB}$ possesses a separating vector, then 
\emph{all}
states of that algebra will be vector states, a \emph{norm} dense set of 
which must therefore be entangled across  $(\alg{R}_{A},\alg{R}_{B})$.  
And since the entangled states of $\alg{R}_{AB}$ are open in the 
weak-$^{*}$ topology, they must be open in the (stronger) norm 
topology too --- so we are dealing with a truly generic set of states. It follows, quite independently of the RS theorem, that
\begin{center}\emph{Generic Result: If $\alg{R}_{A}$ and 
$\alg{R}_{B}$ are commuting nonabelian von Neumann algebras either of 
which possesses 
a cyclic vector, and $\alg{R}_{AB}$ possesses a separating vector, 
then the generic state of $\alg{R}_{AB}$ will be entangled across 
$(\alg{R}_{A},\alg{R}_{B})$.}\end{center}
\noindent  
The role that the RS theorem plays is to guarantee that the antecedent conditions of 
this Generic Result are satisfied whenever we consider spacelike separated 
regions (and corresponding algebras) satisfying $(O_{A}\cup 
O_{B})'\not=\emptyset$.  This is a very weak requirement, which is 
satisfied, for example, when we assume both regions are bounded in 
spacetime. In that case, in order to be \emph{certain} that her 
local 
operation in $O_{A}$ (pure or mixed) produced a disentangled state, 
Alice would need the extraordinary ability to distinguish 
the state of $\alg{A}_{AB}$  which results from her operation from the 
generic set states of $\alg{A}_{AB}$  
that are entangled!   

Finally, while we noted in our introduction the irony 
that limitations on disentanglement arise precisely when one considers 
\emph{relativistic} quantum theory, the practical limitations we have 
just identified --- as opposed to the \emph{intrinsic} limits on disentanglement which are the 
subject of the next section --- are not characteristic of AQFT alone.  
In particular, the existence of locally cyclic states does not depend on field 
theory.  As we have seen, both the $A$ and $B$ subalgebras of 
$\alg{B}(\hil{H}_{A}\otimes \hil{H}_{B})$ possess a cyclic vector 
just in case $\dim\hil{H}_{A}=\dim\hil{H}_{B}$.  Indeed, operator 
algebraists so often find themselves dealing with von Neumann algebras 
that, together with their commutants, possess a cyclic vector, that such algebras 
are said by them to be in `standard form'.  So we should not 
think that local cyclicity is somehow 
peculiar to the states of local quantum 
fields.

Neither is it the case that our Generic Result above finds its only 
application in quantum \emph{field} theory.  For example, consider the infinite-by-infinite state space 
$\hil{H}_{A}\otimes\hil{H}_{B}$ of any two nonrelativistic  
particles, ignoring their spin degrees of freedom.  Take the tensor 
product with a third auxiliary infinite-dimensional Hilbert space 
$\hil{H}_{A}\otimes\hil{H}_{B}\otimes\hil{H}_{C}$.  Then obviously 
$\infty=\dim\hil{H}_{C}\geq\dim(\hil{H}_{A}\otimes\hil{H}_{B})=\infty$, 
whence the $C$ subalgebra
possesses a cyclic vector, which is therefore 
separating for the $A+B$ algebra.  On the same dimensional grounds, both the $A$ and $B$ subalgebras 
possess cyclic vectors of their own.  So our Generic Result applies 
immediately yielding the conclusion that
a typical state of $A+B$ will be 
entangled (cf. Clifton and Halvorson, 2000).

Nor should we think of local cyclicity or the applicability of our  
Generic Result as peculiar to standard  
\emph{local} quantum field theory.
 After noting that the local cyclicity of the vacuum in AQFT was a 
``great, counterintuitive, surprise'' (p. 4) when it was first proved,  Fleming (1999) 
proposes, instead, to build up local algebras 
associated with bounded open spatial sets within hyperplanes from 
raising and lowering operators associated with nonlocal 
Newton-Wigner position eigenstates --- a proposal that goes back at 
least as far as Segal (1964).  Fleming then observes, as did Segal 
(1964, p. 143), that the 
resulting vacuum state will \emph{not} be entangled nor cyclic for 
any such local algebra.  Nevertheless, 
as Segal points out, each Segal-Fleming local 
algebra will be isomorphic to the algebra $\alg{B}(\hil{H})$ of all bounded operators on an 
\emph{infinite}-dimensional Hilbert space $\hil{H}$, and algebras associated with 
spacelike-separated regions in 
the same hyperplane commute.  It follows that if we take 
any two spacelike separated bounded open regions $O_{A}$ and $O_{B}$ lying in the same 
hyperplane, $[\alg{A}(O_{A})\cup \alg{A}(O_{B})]''$ will be
 naturally isomorphic to 
 $\alg{B}(\hil{H}_{A})\otimes\alg{B}(\hil{H}_{B})$ (Horuzhy 1988, 
 Lemma 1.3.28), and the result 
of the previous paragraph applies.  So 
Fleming's `victory' over the RS theorem of standard local quantum 
field theory rings hollow. Even though the Newton-Wigner 
vacuum is not itself entangled or locally cyclic across the regions 
$(O_{A},O_{B})$, it will be 
indistinguishable from globally pure states of the Newton-Wigner 
field that are!  (For further discussion of the Segal-Fleming 
approach to quantum fields, see Halvorson (2000).) 

On the other hand, generic entanglement is certainly not to be expected in every 
quantum-theoretic context.  For example, if we ignore external degrees of freedom, and 
just consider the spins of two particles with joint state space 
$\hil{H}_{A}\otimes\hil{H}_{B}$, where both 
spaces are nontrivial and \emph{finite}-dimensional,  
then the Generic Result no longer applies.  Taking the product with a 
third auxiliary Hilbert space $H_{C}$ does not work, because in order 
for the $A+B$ subalgebra to have a separating vector we would need 
$\dim H_{C}\geq \dim H_{A}\dim H_{B}$, but for either the $A$ or $B$ subalgebras 
to possess a cyclic vector we would \emph{also} need that either 
$\dim H_{A}\geq \dim H_{B}\dim H_{C}$ or 
$\dim H_{B}\geq \dim H_{A}\dim H_{C}$ --- both of which contradict the 
fact  $\hil{H}_{A}$ and $\hil{H}_{B}$ are nontrivial and 
finite-dimensional.    
(In fact, it can be shown 
that the spins of any pair of particles are \emph{not} generically 
entangled, unless of course we ignore their mixed spin states; see Clifton and 
Halvorson, 2000 for further discussion.)  
The point is that while the conditions for 
generic entanglement may or 
may not obtain in \emph{any} quantum-theoretical context --- depending 
on the observables and dimensions of the state spaces involved --- the 
beauty of the RS theorem is that it allows us to deduce that generic 
entanglement between bounded open spacetime regions 
\emph{must} obtain just by making some very general and natural 
assumptions about what should count as a 
physically reasonable relativistic quantum field theory.

\section*{4. Type III von Neumann Algebras and \\  Intrinsic Entanglement}

 Though it is not known to follow from the general axioms of AQFT 
 (cf. Kadison, 
 1963), all known concrete 
 models of the axioms are such that the local algebras associated with 
 bounded open regions in $M$ are type 
 III factors (Horuzy, 1988, pgs. 29, 35; Haag, 1992, Sec. V.6).  We start 
  by reviewing 
 what precisely is meant by the designation `type III factor'.
 
 A von Neumann algebra $\alg{R}$ is a factor just in case its center 
 $\alg{R}\cap\alg{R}'$ consists only of multiples of the identity.  
 It is easy to verify that this is equivalent to 
 $(\alg{R}\cup\alg{R}')''=\alg{B}(\hil{H})$, thus $\alg{R}$ induces a 
 `factorization' of
 the total Hilbert space algebra $\alg{B}(\hil{H})$ into two subalgebras which together 
 generate that algebra.

  To understand what `type III' means, a few further definitions need 
  to be absorbed.  A partial isometry $V$ is an operator on a Hilbert space $\hil{H}$ that maps 
 some particular closed subspace $C\subseteq\hil{H}$ isometrically onto 
 another closed subspace $C'\subseteq\hil{H}$, and maps $C^{\perp}$ to 
 zero.  (Think of $V$ as a `hybrid' unitary/projection operator.)  
 Given the set of projections in a von Neumann algebra 
 $\alg{R}$, we can define the following equivalence relation on 
 this set: $P\sim Q$ just in case there is a partial isometry 
 $V\in\alg{R}$ that maps the range 
 of $P$ onto the range of $Q$.  (It is important to notice that this 
 definition of equivalence is relative to the particular von Neumann algebra 
 $\alg{R}$ that the projections are considered to be members of.) For example, any two infinite-dimensional 
 projections in $\alg{B}(\hil{H})$ are equivalent (when $\hil{H}$ is 
 separable), including projections 
 one of whose range is properly contained in the other (cf. KR 1997, 
 Cor. 6.3.5).  A nonzero projection $P\in\alg{R}$ is called 
 abelian if the von Neumann algebra $P\alg{R}P$ acting on the 
 subspace $P\hil{H}$ (with identity $P$) is abelian.  One can 
 show that the abelian projections in a factor $\alg{R}$ are exactly 
 the atoms in its projection lattice (KR 1997, Prop. 
 6.4.2).  For example, the atoms of the projection lattice of 
 $\alg{B}(\hil{H})$ are all 
 its one-dimensional projections, and they are all (trivially) abelian, 
 whereas it is clear that higher-dimensional projections are not. Finally, a projection $P\in\alg{R}$ is 
 called infinite (relative to $\alg{R}$!) when it is equivalent to 
 another projection $Q\in\alg{R}$ such that $Q<P$, i.e., Q projects onto a 
 proper subspace of the range of $P$.  One can also show that any 
 abelian projection in a von Neumann algebra must be \emph{finite}, 
 i.e., not infinite 
 (KR 1997, Prop. 6.4.2).  
 
 A type I von Neumann factor is now defined as one that possesses an 
 abelian projection.  For example, $\alg{B}(\hil{H})$ for any Hilbert 
 space $\hil{H}$ is always type I, and, indeed, every type I factor arises 
 as the algebra of all bounded operators on some Hilbert space (KR 
 1997, Thm 6.6.1).  On the other hand, a factor is type III 
 if all its nonzero projections are infinite and equivalent.  In particular, this 
 entails that the algebra itself is not abelian, nor could it even possess an
 abelian projection --- which would have to be finite.  
 And since a type III factor 
 contains no abelian projections, its projection lattice cannot have any 
 atoms.  Another fact about type III algebras is that they \emph{always} 
 possess a vector that is both cyclic and separating (Sakai, 1971, Cor. 
 2.9.28).   Therefore we know that type III algebras will always
 possess a dense set of cyclic vectors, and that all their states will be 
 vector states.  \emph{Notwithstanding this}, type III algebras possess 
 \emph{no} pure states, as a consequence of the fact that they lack 
 atoms.  
 
 To get some feeling for why this is the case --- and for the general 
 connection between the failure of the projection lattice of an 
 algebra to possess atoms and its failure to 
 possess pure states --- let $\alg{R}$ be any non-atomic von Neumann algebra 
 possessing a separating vector (so all of its states are vector states), and let $\rho_{x}$ be any state of 
 $\alg{R}$.  We shall need two further definitions.  The support projection, 
 $S_{x}$, of $\rho_{x}$ in 
 $\alg{R}$ is defined to be the meet of all projections $P\in\alg{R}$ 
 such that $\rho_{x}(P)=1$.  (So $S_{x}$ is the smallest projection 
 in $\alg{R}$ that $\rho_{x}$ `makes true'.)  The 
 left-ideal, $I_{x}$, of $\rho_{x}$ in $\alg{R}$ is defined to 
 be the set of all $A\in\alg{R}$ such that $\rho_{x}(A^{*}A)=0$.   
 Now since $S_{x}$ 
 is not an atom, there is some nonzero $P\in\alg{R}$ such that $P<S_{x}$. 
 Choose any vector $y$ in the range of $P$ (noting it follows that 
 $S_{y}\leq P$).  We shall first show that $I_{x}$ is a proper subset of 
 $I_{y}$.   So let $A\in	I_{x}$.	 Clearly this is equivalent to saying 
 that $Ax=0$, or that $x$ lies in the range of $N(A)$, the projection onto the null-space of $A$.  
 $N(A)$ itself lies $\alg{R}$ (KR 1997, Lemma 5.1.5 and Prop. 2.5.13), 
 thus, $\rho_{x}(N(A))=1$, and accordingly $S_{x}\leq N(A)$.   
 But since $S_{y}\leq P<S_{x}$, we also have	$\rho_{y}(N(A))=1$.  Thus,	
 $y$ too lies in the range of $N(A)$, i.e., $Ay=0$, and therefore $A\in 
 I_{y}$. To see that the inclusion $I_{x}\subset I_{y}$ is 
 proper, note that since $(y,S_{y}y)=1$, $(y,[I-S_{y}]^{2}y)=0$, and thus 
 $I-S_{y}\in I_{y}$.  However, certainly $I-S_{y}\not\in I_{x}$, for the 
 contrary would entail that 
 $(x,S_{y}x)=1$, in other words, $S_{x}\leq S_{y}\leq P<S_{x}$ --- a 
 contradiction.  We can now see, finally, that 
 $\rho_{x}$ cannot be pure.  For, quite generally, the pure states of a von Neumann 
 algebra $\alg{R}$ determine \emph{maximal} left-ideals in 
 $\alg{R}$ (KR 1997, Thm. 10.2.10), yet we have just shown, under the 
 assumption that $\alg{R}$ is non-atomic, that $I_{x}\subset I_{y}$. 
 
 The fact that every state of a type III algebra $\alg{R}$ is 
 mixed throws an entirely new wrench into the works of the ignorance 
 interpretation of mixtures.  (To our knowledge, Van Aken (1985) is the only 
 philosopher of quantum theory to 
 have noticed this.)   Not only is  there no preferred way
  to pick out components of a mixture, but the components of 
 states of $\alg{R}$ will always \emph{themselves} be mixtures.  Thus, 
 it is impossible to understand the preparation of such a mixture
 in terms of mixing pure states --- 
 the states of $\alg{R}$ are always irreducibly or \emph{intrinsically} 
 mixed.  
 Note, however, that while the states of type III factors fit this description, so 
 do the states of certain \emph{abelian} von Neumann algebras.  For example, the  
 `multiplication' algebra $\alg{M}\subseteq \alg{B}(L_{2}(\mathbb{R}))$ of 
 all bounded functions of the position operator for a single 
 particle lacks atomic projections because position has no 
 eigenvectors.  Moreover, all 
 the states of $\alg{M}$ are vector states, because any state vector 
 that corresponds to a wavefunction 
 whose support is the whole of $\mathbb{R}$ is separating for 
 $\alg{M}$.  Thus 
 the previous paragraph's argument applies equally well to $\alg{M}$.  
 
 Of course
  no properly \emph{quantum} system has an abelian algebra of 
 observables, and, as we have already noted, systems with abelian algebras 
 are never entangled with other systems.  This makes the failure of a 
 type III factor $\alg{R}$ to have pure states importantly different from that 
 failure in the case of an abelian algebra. 
 Because $\alg{R}$ is \emph{non}abelian, and taking the 
 commutant preserves type (KR 1997, Thm. 9.1.3) so that $\alg{R}'$ 
 will also be nonabelian, one suspects
 that any pure 
 state of $(\alg{R}\cup\alg{R}')''=\alg{B}(\hil{H})$ --- which must 
 restrict to an intrinsically mixed state on both subalgebras 
 $\alg{R}$ and $\alg{R}'$ --- has to be 
 \emph{intrinsically entangled} across $(\alg{R},\alg{R}')$.  And that intuition is exactly 
 right; indeed, one can show that there 
 are not even any \emph{product} states across $(\alg{R},\alg{R}')$ 
 (Summers 1990, p. 
 213).  
 And, of course, if there are no unentangled states across 
 $(\alg{R},\alg{R}')$, then the infamous distinction, some have argued is 
 important to preserve, between so-called `improper' mixtures 
 that arise by restricting an entangled state to a subsystem, and 
 `proper' mixtures that do not, becomes \emph{irrelevant}.  
 
 Even more interesting is the fact that in all known models of AQFT, the local algebras 
  are `type III$_{1}$'.  It would take us too far 
 afield to explain the standard sub-classification of factors presupposed by 
 the subscript `$1$'.  We wish only to draw attention to 
 an equivalent characterization of type III$_{1}$ algebras 
 established by Connes and St\o rmer 
 (1978, Cor. 6): A factor $\alg{R}$ acting standardly on a 
 (separable) Hilbert space is type III$_{1}$ just in case 
 for \emph{any two} states $\rho,\omega$ of $\alg{B}(\hil{H})$, and any 
 $\epsilon>0$, there are unitary operators 
 $U\in\alg{R}$, $U'\in\alg{R}'$ such 
 that $\|\rho-\omega^{UU'}\|<\epsilon$.  Notice that this result 
 immediately implies that there can be no unentangled states across 
 $(\alg{R},\alg{R}')$; for, if some $\omega$ were not entangled, it 
 would be impossible to act on this state with local unitary 
 operations in $\alg{R}$ and $\alg{R}'$ and get arbitrarily close to 
 the states that \emph{are} entangled across $(\alg{R},\alg{R}')$.  
 Furthermore --- and this is the interesting fact --- the Connes-St\o rmer characterization 
 immediately implies the  
 impossibility of distinguishing in any reasonable way between the different degrees of 
 entanglement that states might have across $(\alg{R},\alg{R}')$.  
 For it is a standard assumption in quantum information theory 
 that 
 all reasonable measures of entanglement must be \emph{invariant} under 
 unitary operations on the separate entangled systems (cf. Vedral 
 \emph{et al}, 1997), and presumably such a measure 
 should assign close 
 degrees of entanglement to states that are close to each other in 
 norm.  In light of the Connes-St\o rmer characterization, imposition 
 of
 both these requirements forces triviality on any 
 proposed measure of entanglement across $(\alg{R},\alg{R}')$.  
 Of course, the standard von Neumann 
 entropy measure we discussed in Section 1. is norm continuous, and, 
 because of the unitary invariance of the trace, this measure is 
 invariant under unitary operations on the component systems.  
 But in the case of a type III factor $\alg{R}$, that measure, as we 
 should expect, is 
 \emph{not} available.  Indeed, the state of a system described by $\alg{R}$ cannot be 
 represented by any density 
 operator \emph{in} $\alg{R}$ because $\alg{R}$  cannot contain compact operators, like 
 density operators, whose spectral projections 
 are all finite!  
  
 The above considerations have particularly strong physical 
 implications when we consider local algebras associated with diamond 
 regions in $M$, i.e., regions given by the intersection of the 
 timelike future of a given spacetime point $p$ with the timelike past of 
 another point in $p$'s future.  
When $\diamondsuit\subseteq M$ is a diamond,  it can be shown in many models of 
AQFT, including for \emph{non}interacting fields, that
 $\alg{A}(\diamondsuit')=\alg{A}(\diamondsuit)'$ (Haag 
 1992, Sec. III.4.2).  
 Thus every global state of the field 
 will be intrinsically entangled across $(\alg{A}(\diamondsuit),\alg{A}(\diamondsuit'))$, and it is 
 never possible to think of the field system in a diamond region $\diamondsuit$ as disentangled from 
 that of its spacelike complement.  Though he does not use the 
 language of entanglement, this is precisely the reason 
 for Haag's remark that field systems are always 
 open.  In particular, Alice would have \emph{no hope 
 whatsoever} of using local operations in $\diamondsuit$ to disentangle that 
 region's state from that of the rest of the world.  
 
 Suppose, however, that Alice has only the more limited goal of 
 disentangling a state of the field across some isolated pair of 
 \emph{strictly} spacelike-separated regions 
 $(O_{A},O_{B})$, i.e., regions which remain spacelike 
 separated when either is displaced by an arbitrarily small amount.  
  It is also known that in many models of AQFT  the local 
 algebras possess the split property: for any bounded open $O\subseteq 
 M$, and any larger region $\tilde{O}$ whose interior contains the 
 closure of $O$, there is a type I factor $\alg{N}$ such that 
 $\alg{A}(O)\subset \alg{N}\subset\alg{A}(\tilde{O})$ (Bucholz 1974, Werner 
 1987).  This implies that the von Neumann algebra generated by a pair 
 of algebras for strictly spacelike separated regions is isomorphic to their tensor 
 product and, as a consequence, that there \emph{are} product states across 
 $(\alg{A}(O_{A}),\alg{A}(O_{B}))$ (cf. Summers 1990, pgs. 
 239-40).  Since, therefore, not every state 
 of $\alg{A}_{AB}$ is entangled, we might hope that whatever the global field 
 state is, 
 Alice could \emph{at least in principle} 
 perform an operation in $O_{A}$ on that state that disentangles it across 
 $(O_{A},O_{B})$.  However, we are now going use the fact that  
 $\alg{A}(O_{A})$ lacks abelian projections  to show that a norm 
 dense set of entangled states of
       $\alg{A}_{AB}$ cannot be disentangled by 
       any pure local operation performed in $\alg{A}(O_{A})$. 
                   
  Let $\rho_{x}$ be any one of the norm dense set of entangled states of
       $\alg{A}_{AB}$  induced by a vector $x\in\hil{H}$ cyclic for 
  $\alg{A}(O_{B})$, and let $K\in\alg{A}(O_{A})$ 
  be an arbitrary Kraus operator.  (Observe that  $\rho_{x}^{K}\not=0$ 
       because $x$ is separating for $\alg{A}(O_{B})'$ --- which includes 
       $\alg{A}(O_{A})$ --- and $K^{*}K\in\alg{A}(O_{A})$ is 
       positive.)  Suppose, for the purposes of extracting a contradiction, that  
       $\omega_{x}^{K}$ is not also entangled.
       Let $Ky$, with $y\in\hil{H}$, 
       be any nonzero vector in the range of $K$.  Then, since $x$ is 
       cyclic for $\alg{A}(O_{B})$, we have, for some sequence  
       $\{B_{i}\}\subseteq\alg{A}(O_{B})$,
       $Ky=K(\lim B_{i}x)=\lim(B_{i}Kx)$, which entails
                                           $\|(\omega_{x}^{K})^{B_{i}/|B_{i}|}-\omega_{Ky}\|\rightarrow 0$.
              Since $\omega_{x}^{K}$ is not entangled across 
       $(\alg{A}(O_{A}),\alg{A}(O_{B}))$, and the local pure 
       operations on $\alg{A}(O_{B})$ given by the Kraus operators 
       $B_{i}/|B_{i}|$ cannot 
       create entanglement, 
       we see that $\omega_{Ky}$ is the norm (hence weak-$^{*}$) limit of a 
       sequence of unentangled states and, as such, is not itself 
       entangled either.  Since $y$ was arbitrary, it 
       follows that every nonzero vector in the range of $K$ induces 
       an unentangled state across $(\alg{A}(O_{A}),\alg{A}(O_{B}))$.  
       Obviously, the same conclusion follows for any nonzero vector in the range of $R(K)$ --- the 
       range projection of $K$ --- since the range of the latter lies 
       dense in that of the former.  
       
       Next, consider the von 
       Neumann algebra
       \begin{equation}
       \alg{C}_{AB}\equiv [R(K)\alg{A}(O_{A})R(K)\cup 
       R(K)\alg{A}(O_{B})R(K)]''
       \end{equation}
       acting on 
       the Hilbert space 
       $R(K)\hil{H}$.
              Since $K\in\alg{A}(O_{A})$, $R(K)\in\alg{A}(O_{A})$ (KR 1997, p. 309), and 
       thus the subalgebra $R(K)\alg{A}(O_{A})R(K)$ cannot be abelian --- on pain of 
       contradicting the fact that $\alg{A}(O_{A})$ has no 
       abelian projections.  And neither is $R(K)\alg{A}(O_{B})R(K)$ abelian.  For 
       since $\alg{A}(O_{B})$ itself is nonabelian, there are $Y_{1},Y_{2}\in\alg{A}(O_{B})$ such 
       that $[Y_{1},Y_{2}]\not=0$.  And because our regions 
       $(O_{A},O_{B})$ are strictly spacelike separated, they have 
       the Schlieder property: $0\not=A\in \alg{A}(O_{A}), 
       0\not=B\in \alg{A}(O_{B})$ implies $AB\not=0$ (Summers 1990, 
       Thm. 6.7).
       Therefore, 
       \begin{equation}
       [R(K)Y_{1}R(K),R(K)Y_{2}R(K)]=[Y_{1},Y_{2}]R(K)\not=0.
       \end{equation}
       So we see that neither algebra occurring in $\alg{C}_{AB}$ is 
       abelian; yet they commute, and so there must be at least one 
       entangled state across those algebras.  
       But this conflicts with the conclusion of the preceding 
       paragraph!  For the vector states of $\alg{C}_{AB}$ are precisely 
       those induced by the vectors in the range of $R(K)$, and 
       we deduced that these all induce unentangled states across 
       $(\alg{A}(O_{A}),\alg{A}(O_{B}))$.  Therefore, by restriction, they all 
       induce unentangled states across the algebra $\alg{C}_{AB}$.  But 
       if none of
       $\alg{C}_{AB}$'s 
       vector states are 
       entangled, it can possess \emph{no} entangled states at all.
       
       The above argument still goes through under the weaker assumption that 
       Alice applies any mixed \emph{projective} operation, i.e., any 
       operation $T$ corresponding to a standard von Neumann 
       measurement associated with a mutually orthogonal set 
       $\{P_{i}\}\in\alg{A}(O_{A})$ of projection operators.  For if we 
        suppose, again for reductio,
        that $\rho_{x}^{T}=\sum_{i}\lambda_{i}\rho_{x}^{P_{i}}$ is 
        not entangled across the regions, then since entanglement 
        cannot be created by a further application to $\rho_{x}^{T}$ of the local projective 
        operation given by (say) $T_{1}(\cdot)=P_{1}(\cdot)P_{1}$, it follows that 
        $(\rho_{x}^{T})^{T_{1}}=(\rho_{x}^{T_{1}\circ T})=\rho_{x}^{P_{1}}$ must again be 
        unentangled, and the above reasoning to a contradiction 
        goes through \emph{mutatis mutandis} with $K=P_{1}$.  This is to be 
        contrasted to the nonrelativistic case we considered in 
        Section 1, where Alice \emph{was} able to disentangle an 
        arbitrary state of $\alg{B}(H_{A}\otimes H_{B})$ by a 
        nonselective projective operation on $A$.  And a moment's reflection 
        will reveal that that was possible precisely because of 
        the availability of abelian projections in the algebra of her 
        subsystem $A$.  
        We have not, of course, shown that the above 
        argument covers \emph{arbitrary} mixing operations Alice 
        might perform in $O_{A}$; in particular, positive-operator 
        valued mixings, where the Kraus 
        operators $\{K_{i}\}$ of a local operation $T$ in $O_{A}$ do not have mutually 
        orthogonal ranges.  However, although it would be interesting to know 
        how far the result could be pushed, we have already expressed our 
        reservations about whether arbitrary mixing operations should count 
        as disentangling when none of the pure operations of which they are 
        composed could possibly produce disentanglement on their own.   
        
                    In summary:
                    \begin{center} \emph{There are many 
                    regions of spacetime within which no local operations can be 
               performed that will disentangle that region's state from that of 
               its spacelike complement, and within which no pure or 
               projective 
               operation on any one of a norm dense set of states can 
               yield 
               disentanglement from the state of any other 
               strictly spacelike-separated region.}\end{center}
\noindent 
Clearly the advantage of 
               the formalism of AQFT is 
               that it allows us to see clearly just how much more 
               deeply entrenched entanglement is in 
               \emph{relativistic} quantum theory.  At the very least, this should 
               serve as a strong note of caution to those who would quickly 
               assert that quantum nonlocality cannot peacefully 
               exist with relativity. 
               
               As far as what becomes of Einsteinian worries about the possibility of 
                doing science in such a deeply entangled world, the split property of local algebras comes to 
                the rescue.   For let us suppose Alice knows nothing more than 
                that she wants to prepare some 
                state $\rho$ on $\alg{A}(O_{A})$ for subsequent 
                testing.  (The following argument is simply an 
                amplification of 
                the reasoning in Werner 1987 and 
                Summers 1990, Thm. 3.13.)   
                Since there is a type I factor $\alg{N}$ satisfying  
                $\alg{A}(O_{A})\subset 
                \alg{N}\subset\alg{A}(\tilde{O}_{A})$ 
                for any super-region $\tilde{O}_{A}$, and $\rho$ is a 
                vector state (when we assume $(O_{A})'\not=\emptyset$), 
                its vector representative defines a 
                state on $\alg{N}$ that extends $\rho$ and is, therefore, represented by some density 
                operator $D_{\rho}$ in the type I algebra $\alg{N}$.  
                Now
                $D_{\rho}$ is an infinite convex 
                combination $\sum_{i}\lambda_{i}P_{i}$ of mutually orthogonal 
                atomic projections 
                in $\alg{N}$ satisfying $\sum_{i} P_{i}=I$ with $\sum_{i} 
                \lambda_{i}=1$.  But each 
                such projection is equivalent, \emph{in the type} III 
                \emph{algebra} $\alg{A}(\tilde{O}_{A})$, to the 
                identity operator.  
                Thus, 
                for each $i$, there is a partial isometry $V_{i}\in\alg{A}(\tilde{O}_{A})$  
                satisfying $V_{i}V_{i}^{*}=P_{i}$ and 
                $V_{i}^{*}V_{i}=I$.  Next, consider the 
                nonselective operation $T$ on $\alg{A}(\tilde{O}_{A})$ given by 
                Kraus operators $K_{i}=\sqrt{\lambda_{i}}V_{i}$, and 
                fix an arbitrary $X\in\alg{A}(O_{A})$.  We claim that $T(X)=\rho(X)I$.  Indeed, because each 
                $P_{i}$ is abelian in $\alg{N}\supseteq\alg{A}(O_{A})$, the operator $P_{i}XP_{i}$ acting 
                on $P_{i}\hil{H}$ can only be some multiple, $c_{i}$, of the 
                identity operator $P_{i}$ on $P_{i}\hil{H}$, and taking the 
                trace of both sides of the equation
                \begin{equation} \label{eq:this} 
                P_{i}XP_{i}=c_{i}P_{i}
                \end{equation}
                 immediately reveals that 
                $c_{i}=\mbox{Tr}(P_{i}X)$.  Moreover, acting on the 
                left of (\ref{eq:this}) with $V_{i}^{*}$ and on the 
                right with $V_{i}$, we obtain 
                $V_{i}^{*}XV_{i}=\mbox{Tr}(P_{i}X)I$, 
                which yields the desired conclusion when multiplied 
                by $\lambda_{i}$ and summed over 
                $i$.   Finally, since $T(X)=\rho(X)I$ for all 
                $X\in\alg{A}(O_{A})$, obviously $\omega^{T}=\rho$ for all 
                initial states $\omega$ of $\alg{A}(O_{A})$.  Thus, once we 
                allow Alice to perform an operation like $T$ that is \emph{approximately} 
                local to $\alg{A}(O_{A})$ (choosing $\tilde{O}_{A}$ to 
                approximate $O_{A}$ as close as we like), she has the 
                freedom to 
                prepare any state of $\alg{A}(O_{A})$ that she pleases.  
                
                Notice that, ironically, testing the theory is actually  
                \emph{easier} here than in 
                nonrelativistic quantum theory!  For we were able to exploit 
                above the type
                III character of  $\alg{A}(\tilde{O}_{A})$ to show that Alice 
                can always prepare her desired state on $\alg{A}(O_{A})$ 
                \emph{nonselectively}, i.e., without ever 
                having to sacrifice any members of her ensemble!  
                Also observe that the result of her preparing 
                operation $T$, because it is local to 
                $\alg{A}(\tilde{O}_{A})$,   
                will always produce a product state across $(O_{A},O_{B})$ 
                when $O_{B}\subseteq (\tilde{O}_{A})'$.  That is, for any initial 
                state $\omega$ across the regions, and all 
                $X\in\alg{A}(O_{A})$ and $Y\in\alg{A}(O_{B})$, we have 
                \begin{equation}
                 \omega^{T}(XY)=\omega(T(X)Y)=\omega(\rho(X)Y)=\rho(X)\omega(Y).
                \end{equation}  
                So 
                as soon as we allow Alice to perform \emph{approximately} local 
                operations on her field system, she \emph{can} isolate it from 
                entanglement with other strictly spacelike-separated 
                field systems, while simultaneously preparing its state 
                as she likes and with 
                relative ease.
               God 
               is subtle, but not malicious.\vspace{.1in}
               
 \noindent \emph{Acknowledgments}---The authors are grateful to Paul 
Busch for helpful discussions, Jeremy Butterfield for helping us to 
clarify our critique of Redhead's discussion of the operational 
implications of cyclicity, and 
Reinhard Werner for filling in for us the argument of Eqn. 
(\ref{eq:werner}).  R. K. C. wishes to thank All Souls College, Oxford 
for support under a Visiting Fellowship.  

     \begin{center} {\bf References} \end{center} 
 
 \noindent Bacciagaluppi, G. (1993), `Separation Theorems and Bell 
 Inequalities in Algebraic Quantum Mechanics',  in P. Busch, P. Lahti, 
 and P. 
Mittelstaedt (eds), \emph{Symposium on the Foundations of Modern 
Physics 1993} (Singapore: World Scientific) pp. 29--37.\vspace{.1in}  
  
  \noindent Bennett, C. H., DiVincenzo, D. P., Fuchs, C. A.,  
Mor, T., Rains, E., Schor, P. W., Smolin, J. A., and Wooters, W. K. 
(1999), 
`Quantum Nonlocality without Entanglement', \emph{Physical Review A} 
\textbf{59}, 1070--1091.\vspace{.1in}

\noindent Borchers, H. J. (1965), `On the Vacuum State in Quantum 
Field Theory.  II.', \emph{Communications in Mathematical Physics} 
\textbf{1}, 57--79.\vspace{.1in}

\noindent Braunstein, S. L., Caves, C. M., Jozsa, R., Linden, N., Popescu, S., 
and Schack, R. (1999), `Separability of Very Noisy Mixed States and 
Implications for NMR Quantum Computing', \emph{Physical Review 
Letters} \textbf{83}, 1054--1057.\vspace{.1in}

\noindent Bucholz, D. (1974), `Product States for Local Algebras', 
\emph{Communications in Mathematical Physics} \textbf{36}, 
287--304.\vspace{.1in}

\noindent Busch, P., Grabowski, M., and 
Lahti, P. J. (1995), \emph{Operational Quantum Physics} (Berlin: Springer).\vspace{.1in}

\noindent Clifton, R. and Halvorson, H. (2000), `Bipartite mixed states of
infinite-dimensional systems are generically nonseparable', 
\emph{Physical Review A} \textbf{61}, 012108.\vspace{.1in}

\noindent Clifton, R., Feldman, D. V., Halvorson, H., Redhead, M. L. 
G., and Wilce, A. (1998), `Superentangled States', \emph{Physical 
Review A} \textbf{58}, 135--145.\vspace{.1in}

\noindent Clifton, R., Halvorson, H., and Kent, A. (2000), `Non-local
correlations are generic in infinite-dimensional bipartite systems', 
forthcoming in \emph{Physical Review A}, April issue.\vspace{.1in}

\noindent Connes, A. and St\o rmer, E. (1978), `Homogeneity of the State Space of Factors of Type 
III$_{1}$', \emph{Journal of Functional Analysis} \textbf{28}, 187-196.\vspace{.1in}

\noindent Dixmier, J. and Mar\'{e}chal, O. (1971),
`Vecteurs totalisateurs d'une alg\`{e}bre de von Neumann',
\emph{Communications in Mathematical Physics} \textbf{22}, 44--50. \vspace{.1in}

\noindent Davies, E. B. (1976), \emph{Quantum Theory of Open Systems}  
(London: Academic Press).\vspace{.1in}

\noindent Einstein, A. (1948), `Quantenmechanik und 
Wirklichkeit', \emph{Dialectica} \textbf{2}, 320--324.\vspace{.1in}

\noindent Fleming, G. (1999), `Reeh-Schlieder Meets Newton-Wigner', 
forthcoming in \emph{PSA 1998 Vol. II} 
(East Lansing, MI: Philosophy of Science Association). \vspace{.1in} 

\noindent Haag, R. (1992), \emph{Local Quantum Physics}, 2nd edition (New York: Springer).\vspace{.1in}

\noindent Haag, R. and Kastler, D. (1964), `An Algebraic Approach 
to Quantum Field Theory', \emph{Journal of Mathematical 
Physics} \textbf{5}, 848--861.\vspace{.1in}

\noindent Halvorson, H. (2000), `Does Relativity imply Nonlocality? Unpacking 
the Reeh-Schlieder Theorem', forthcoming.\vspace{.1in}

\noindent Halvorson, H. and Clifton. R. (2000), `Generic Bell 
Correlation between Arbitrary Local Algebras in Quantum Field 
Theory', forthcoming in \emph{Journal of Mathematical Physics} April 
issue.\vspace{.1in}

\noindent Horuzhy, S. S. (1988),
\emph{Introduction to Algebraic Quantum Field Theory} (Dordrecht:
Kluwer).\vspace{.1in}

\noindent Howard, D. (1989), `Holism, Separability, and the 
Metaphysical Implications of the Bell Experiments', in J. T. Cushing 
and E. McMullin (eds.), \emph{The Philosophical Consequences of Bell's 
Theorem} (Notre Dame: Notre Dame University Press) pp. 224--253.\vspace{.1in}

\noindent Kadison, R. (1963),
`Remarks on the type of von Neumann algebras of local observables in
quantum field theory', \emph{Journal of Mathematical Physics} \textbf{4}, 1511--1516.\vspace{.1in}

\noindent Kadison, R. and Ringrose, J. (1997), \emph{Fundamentals of the Theory
of Operator Algebras} (Providence, R. I.: American Mathematical 
Society). \vspace{.1in}

\noindent Kraus, K. (1983), \emph{States, Effects, and Operations} 
(Berlin: Springer).\vspace{.1in}

\noindent Laflamme, R. (1998), Review of `Separability of Very Noisy Mixed States and 
Implications for NMR Quantum Computing' by Braunstein \emph{et al} 
(1998), 
in \emph{Quick Reviews in Quantum Computation and Information}, 
http://quantum-computing.lanl.gov/qcreviews/qc/.\vspace{.1in}

\noindent L\"{u}ders, G. (1951), `\"{U}ber die Zustands\"{a}nderung durch den 
Messprozess', \emph{Annalen der Physik} \textbf{8}, 322--328.\vspace{.1in}

\noindent Mor, T. (1998),
`On the Disentanglement of States', 
{\tt quant-ph/9812020}.\vspace{.1in}

\noindent Mor, T. and Terno, D. R. (1999),
`Sufficient Conditions for Disentanglement', 
{\tt quant-ph/9907036}.\vspace{.1in}

\noindent Popescu, S. and Rohrlich, D. (1997), `Thermodynamics and the Measure of 
Entanglement', \emph{Physical Review A} \textbf{56}, R3319--R3321.\vspace{.1in}

\noindent Redhead, M. L. G. (1995), `More Ado about Nothing',
  \emph{Foundations of Physics} \textbf{25}, 123--137.\vspace{.1in}
  
  \noindent Reeh, H. and Schlieder, S. (1961), `Bemerkungen zur 
  unitaraquivalenz von Lorentzinvarianten Feldern', \emph{Nuovo 
  Cimento} \textbf{22}, 1051--1068.\vspace{.1in}
  
\noindent Sakai, S. (1971), \emph{$C^{*}$-Algebras and $W^{*}$-algebras} 
(Berlin: Springer-Verlag).\vspace{.1in}

\noindent Schroer, B. (1998), \emph{A Course on: `Modular Localization and 
Nonperturbative Local Quantum Physics'}, {\tt hep-th/9805093}.\vspace{.1in}

\noindent Schroer, B. (1999), `Basic Quantum Theory and Measurement 
from the Viewpoint of Local Quantum Physics', {\tt quant-ph/9904072}.\vspace{.1in}

\noindent Segal, I. E. (1964), `Quantum Fields and Analysis in the 
Solution Manifolds of Differential Equations', in W. T. Martin and I. 
E. Segal (eds), \emph{Proceedings of a Conference on the Theory and 
Applications of Analysis in Function Space} (Boston: MIT Press) Ch. 8.\vspace{.1in}

\noindent Segal, I. E. and Goodman, R. W. (1965), `Anti-Locality 
of Certain Lorentz-Invariant Operators', \emph{Journal of Mathematics and 
Mechanics} \textbf{14}, 
629--638. \vspace{.1in}

\noindent Streater, R. F. and Wightman, A. S. (1989), 
\emph{PCT, Spin and Statistics, and All That} (New York: Addison-Wesley).\vspace{.1in}

\noindent Summers, S. J. (1990), `On the independence of local
  algebras in quantum field theory', \emph{Reviews of Mathematical 
  Physics} \textbf{2}, 201--247.\vspace{.1in}

\noindent Summers, S. J. and Werner, R. F. (1988), `Maximal violation of
Bell's inequalities for algebras of observables in tangent spacetime
regions',  Annales Institut Henri Poincar{\'e} \textbf{49}, 215--243.

\noindent Summers, S. J., and Werner, R. F. (1995), `On Bell's
inequalities and algebraic invariants', \emph{Letters in Mathematical 
Physics} \textbf{33},
321--334.\vspace{.1in}

\noindent Van Aken, J. 
(1985), 
`Analysis of Quantum Probability Theory', \emph{Journal of 
Philosophical Logic} 
\textbf{14}, 267--296.\vspace{.1in}

\noindent Van Fraassen, B. C. (1991), \emph{Quantum Mechanics: An 
Empiricist View} (Oxford: Clarendon Press).\vspace{.1in}

\noindent Vedral, V., Plenio, M. B., Rippin, M. A., and Knight, P. L. 
(1997), 
`Quantifying Entanglement', \emph{Physical Review Letters} \textbf{78}, 
2275--2279.\vspace{.1in}

\noindent Werner, R. F. (1987), `Local Preparability of States and the 
Split Property', \emph{Letters in Mathematical Physics} \textbf{13}, 325--329.\vspace{.1in}

\noindent Werner, R. F. (1989), `Quantum States with
  Einstein-Podolsky-Rosen correlations admitting a hidden-variable
  model', \emph{Physical Review A} \textbf{40}, 4277--4281.\vspace{.1in}

\end{document}